\documentclass[pre,aps,twocolumn,showpacs]{revtex4}

\usepackage{dcolumn}
\usepackage{amsmath}
\usepackage{amssymb,amsfonts,latexsym}
\usepackage{bm}
\usepackage[mathcal]{euscript}
\usepackage{graphicx}
\usepackage{epsfig}
\usepackage{color}

\definecolor{blue}{rgb}{0,0,1}
\definecolor{green}{rgb}{0,0.5,0}
\definecolor{red}{rgb}{1,0,0}
\definecolor{pink}{rgb}{0.9,0.3,0.7}
\definecolor{azur}{rgb}{0,0.5,0.5}
\definecolor{orange}{rgb}{1,0.5,0.2}
\definecolor{brown}{rgb}{0.5,0,0}

\newcommand{\be}{\begin{equation}}
\newcommand{\ee}{\end{equation}}
\newcommand{\ben}{\begin{equation*}}
\newcommand{\een}{\end{equation*}}
\newcommand{\bea}{\begin{eqnarray}}
\newcommand{\eea}{\end{eqnarray}}

\newcommand \ve {\varepsilon}

\begin{document}
\graphicspath{{figures/}}

\title{Subcritical transition to turbulence: what we can learn from the physics of glasses}

\author{Olivier Dauchot$^1$ and Eric Bertin$^2$}

\affiliation{$^1$ EC2M, ESPCI-ParisTech, UMR Gulliver 7083 CNRS, 75005 Paris, France\\
$^2$ Universit\'e de Lyon, Laboratoire de Physique, ENS Lyon, CNRS, 46 All\'ee d'Italie, 69007 Lyon, France}

\date{\today}

\begin{abstract}
In this note, we discuss possible analogies between the subcritical transition
to turbulence in shear flows and the glass transition in supercooled liquids.
We briefly review recent experimental and numerical results, as well as theoretical
proposals, and compare the difficulties
arising in assessing the divergence of the turbulence lifetime in subcritical shear flow
with that encountered for the relaxation time in the study of the glass transition.
In order to go beyond the purely methodological similarities, we further elaborate
on this analogy and propose a simple model for the transition to turbulence,
inspired by the Random Energy Model (a standard model for the glass transition),
with the aim to possibly foster yet unexplored directions of research in subcritical shear flows.
\end{abstract}


\maketitle

\section{Introduction}
Statistical physics has devoted a lot of effort to the study of fully developed turbulence,
but much less to that of the transition to turbulence~\cite{Frisch:318247}, which occurs when the
Reynolds number, the ratio of the advection time to the viscous time, is increased. 
The transition is commonly observed in flow regimes lacking linear instability
and is referred to as globally subcritical~\cite{Joseph:1976twa,Grossmann:2000zz, Dauchot:1997wta}.

The plane Couette flow, driven by two plates moving parallel to each other in opposite
directions is linearly stable at all Reynolds number and, as such, is the epitome of
globally subcritical transitions~\cite{Romanov73}.
Other flows usually transit to turbulence before linear instability sets in.
These include the circular Poiseuille flow (cPf) and the plane Poiseuille flow (pPf),
which are driven by a pressure gradient respectively along a circular pipe or
between two parallel plates, as well as the counter-rotating
Taylor Couette flow (TCf), driven by two concentric cylinders rotating in opposite
directions.
In all these cases, the transition is particularly delicate to understand owing to
its abrupt character. A complex spatio-temporal dynamics is observed, involving
in particular the nucleation and the growth or decay of turbulent
domains called 'puffs' (pPf) or 'spots' (pCf) --see, e.g.,~\cite{Darbyshire:2006hsa,Peixinho:2008vb}
for cPf,~\cite{Tillmark92,PhysRevLett.69.2511,Dauchot:1995tha} for pCf,
~\cite{Carlson82} for pPf, and~\cite{Coles65, Coles67} for TCf.

A recent surge of interest has been motivated by the audacious proposal
that shear flow turbulence could remain transient up to arbitrarily large
Reynolds number, opening ways towards a better control of such turbulent
regimes~\cite{Hof:2006fk}.
This proposal was motivated by experimental and numerical observations 
in cPf~\cite{Hof:2006fk} and pCf~\cite{TobiasMSchneider:2010gv} regarding the statistics of turbulent lifetimes, which contradicted those previously obtained in cPf~\cite{Faist04:JFM213501,Peixinho:2006kma}
and pCf~\cite{Bottin:1998ve,bottinEPJ98}.
These contradictory results, together with the experimental discovery of a spectacular
long wavelength periodic organization of the laminar-turbulent coexistence in pCf and
TCf~\cite{Prigent:2002fqa,Prigent:2003uma}, has motivated further experiments in
TCf~\cite{PhysRevE.81.025301} and cPf~\cite{Avila:2011up,Samanta:2011va},
the development of various models~\cite{Eckhardt:2008gd,MannevilleEPJ07,Gibson:2008ec,Manneville:2009if,PhysRevE.84.016309} and an impressive number of numerical studies~\cite{PhysRevLett.94.014502,PhysRevLett.98.014501,Schneider:2008ig,Duguet:2008bs,Tuckerman:2008gw,Duguet:2009bd,Schneider:2010id,Duguet:2010dv,Moxey:2010es,Manneville:2010hd,Avila:2011up,Philip:2011ja,Manneville:2012uta}.
As a result, some comprehension of the mechanisms at play in the coexistence dynamics,
as well as a better knowledge of the organization of phase space, involving
many unstable solutions of the Navier-Stokes equation, has been gained.

Interestingly, the presence in phase space of many unstable solutions
and the existence of finite, yet extremely large, relaxation times,
are reminiscent of the physics of glasses
(see, e.g.,~\cite{struik1978physical,PhysRevLett.85.5356,berthier2011theoretical}).
In particular, whether the structural relaxation times of a glass really diverges at a given
finite temperature or remains very large but finite at any positive temperature
is an important question --related to the existence of a genuine phase transition to
an ideal glass state-- that remains largely open \cite{PhysRevE.53.751}.
However, the intense activity related to this specific issue has triggered along the way
different (and perhaps even more interesting) questions, driving
the field of glasses towards important conceptual progresses~\cite{berthier2011theoretical}.

In this note, we explore the analogy between the subcritical transition
to turbulence and the glass transition from several viewpoints.
After a concise review of the major results on the transition
to turbulence, we discuss the limitations of fitting procedures
in assessing the divergence of the turbulence lifetime, drawing inspiration
from similar discussions in the glass literature (Sec.~\ref{sec-turb-lifetime}).
Then, we briefly review the theoretical scenarios and models that have been proposed
to describe the subcritical transition to turbulence,
and we tentatively discuss the analogy with glasses at a conceptual level
(Sec.~\ref{sec-rev-dyn}).
The understanding of the glass transition has greatly
benefited from the study of oversimplified models like the 
Random Energy Model~\cite{PhysRevLett.45.79,PhysRevB.24.2613},
which describes the statistical behavior of a system evolving in a random
energy landscape.
In this spirit, we try to transpose the Random Energy Model, keeping in mind both its strengths
and weaknesses, to the modeling of the subcritical transition
to turbulence, in order to possibly gain insight into the statistical
mechanisms at play in this transition (Sec.~\ref{sec-model}).
As a result, we obtain an estimate of the turbulence lifetime as a function
of the Reynolds number close to the transition, an estimate which qualitatively agrees
amazingly well with the observed phenomenology.

Clearly, this qualitative agreement does not in itself prove the analogy to be specifically deep, but
it suggests that it deserves to be further explored.
More generally, we hope, in the spirit of Pomeau's seminal paper~\cite{Pomeau86},
that the analogy presented here could foster contributions from the statistical physics community to the
old standing problem of the transition to turbulence, taking advantage of
recently developed concepts in the statistical physics of disordered
systems. Conversely, the development of techniques such as Particle Image
Velocimetry, and the exponential increase of the numerical capacities
could help in validating or invalidating the assumptions made on some
properties of turbulence in following the present analogy.

\section{Turbulence lifetime}
\label{sec-turb-lifetime}

\subsection{A review of experimental and numerical observations}

\begin{figure}[t!] 
\center
\vspace{-0.0cm}
\includegraphics[width = 0.95\columnwidth,trim = 0mm 0mm 0mm 0mm, clip]{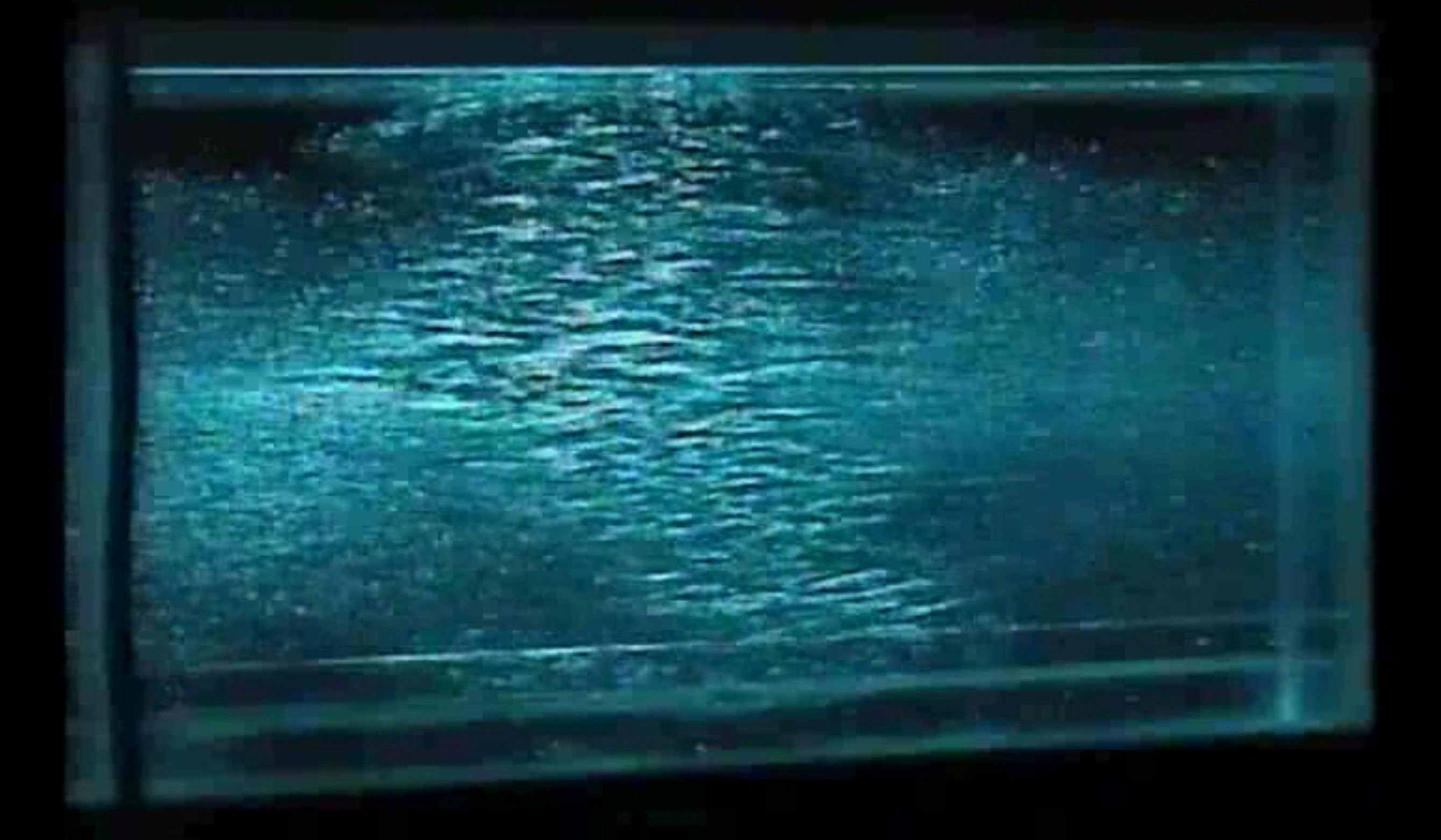}
\includegraphics[width = 0.95\columnwidth,trim = 0mm 0mm 0mm 0mm, clip]{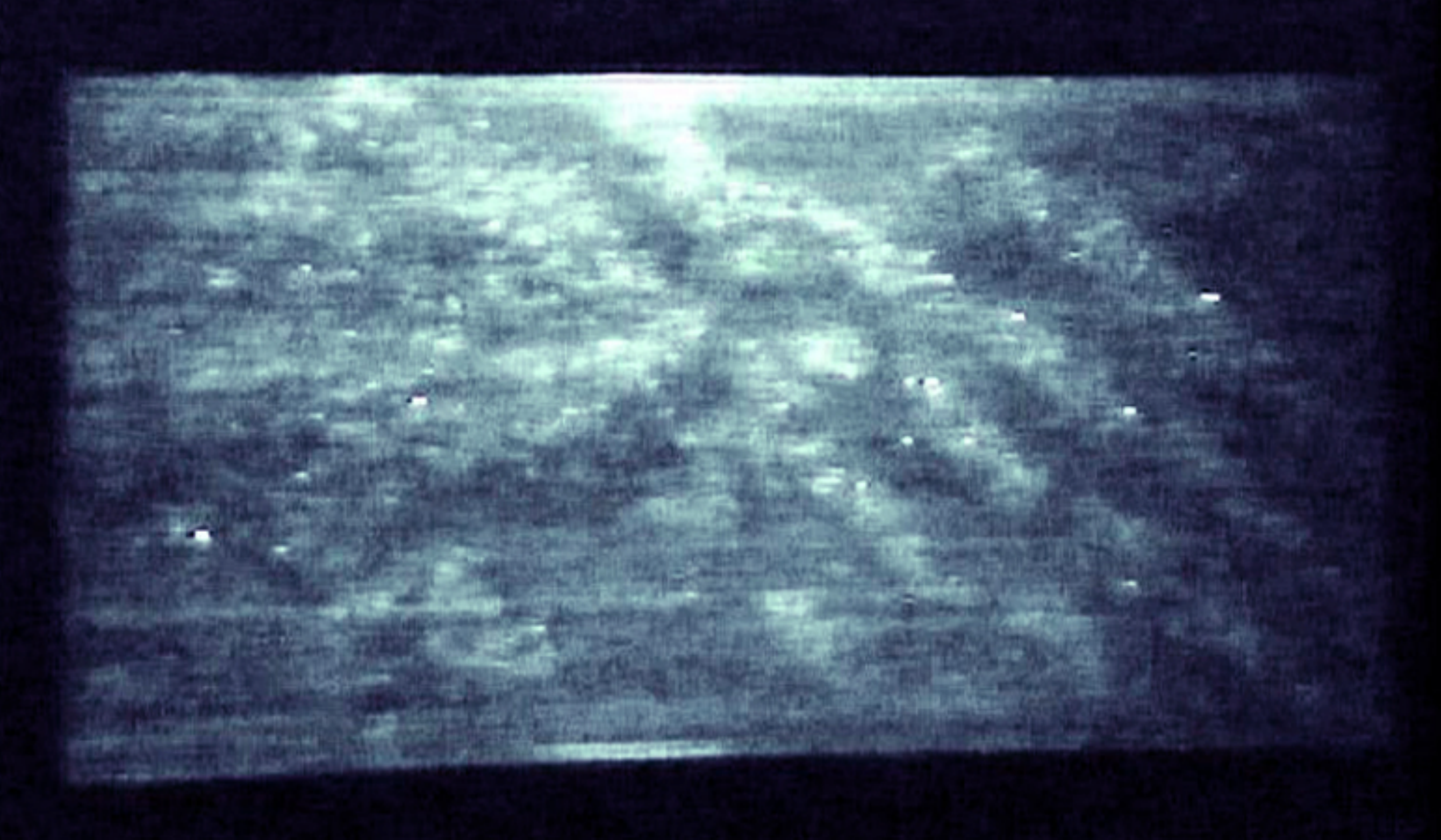}
\caption{Typical snapshots of the laminar-turbulence coexistence at intermediate
Reynolds number in plane Couette flow. Top: turbulent spot in a small aspect ratio setup,
created by a localized perturbation. Bottom: Large scale coexistence in a large aspect ratio
setup, following a quench from high Reynolds number.}
\label{fig:snapshot}
\vspace{-0.5cm}
\end{figure}

A standard characterization of the subcritical transition
to turbulence is the determination of the average turbulence lifetime, following either
a perturbation or a quench, as a function of Reynolds number.
We thus start by briefly reviewing the experiments and direct numerical simulations
reporting the increase of the turbulence lifetime when the Reynolds number
is increased. 
To our knowledge, the first systematic measurement of turbulence lifetimes
was conducted in the pCf~\cite{Bottin:1998ve,bottinEPJ98}.
Two different kinds of experiments
have been performed, differing by the way the initial condition is prepared.
In what we shall call type-A experiments, the Reynolds number, $R$, is set to its
value of interest and the laminar flow is disturbed locally at the initial time. 
In type-B experiments, a turbulent flow is prepared at high $R$, and quenched
at the initial time down to the $R$ value of interest. In both cases, one monitors 
the evolution of the turbulent fraction $f_T(t)$, which characterizes the coexistence
dynamics of laminar and turbulent domains (see Fig~\ref{fig:snapshot}).
For $R>R_g$, $f_T(t)$ fluctuates around some average value, which remains
finite on experimental timescales. For $R<R_u$, $f_T(t)$ relaxes towards zero, 
without displaying any long transient regime. In between, for $R_u<R<R_g$, $f_T(t)$
exhibits a first rapid decay, followed by a long transient quasi-steady regime,
before a large fluctuation sets it to zero. The lifetime  of these transients
are exponentially distributed and the average value $\tau$
was reported to diverge like $(R_g-R)^{-1}$.

The cPf was later investigated in various ways.
In~\cite{Peixinho:2006kma,Darbyshire:2006hsa} a puff 
was generated inside a constant flow rate pipe flow by introducing a short
duration perturbation. Then R was reduced and the subsequent evolution of
the puff was monitored as it progressed downstream. The probability of
observing a localized disturbed region of flow as a function of distance
downstream is exponential and the time required for half of the initial
states to decay, $\tau_{1/2}$, was reported  to diverge like $(R_c-R)^{-1}$,
in agreement with the observations made in the pCf. Other protocols lead
to the same conclusions~\cite{Peixinho:2006kma,Darbyshire:2006hsa}.

However, these results were challenged later by another experimental
study~\cite{Hof:2006fk}. In a pressure driven flow through a very long pipe,
the authors could record much longer dimensionless observation times.
They could determine the probability to be turbulent after a time period
given by the distance between the perturbation location and the outlet,
as a function of flow rate. For short times, the data are within the error bars
of~\cite{Peixinho:2006kma,Darbyshire:2006hsa} but for longer times they
deviate from the divergent behavior reported above and are better represented
by an exponential variation: $\tau=\exp(aR+b)$, without singularity (here and in what follows,
$a$ and $b$ denote generic fit parameters).
Finally in a recent experimental study of turbulence in pipe flow spanning
height orders of magnitude in time, drastically extending all previous
investigations, it was claimed that the turbulent state remains transient,
with a mean lifetime, which depends super-exponentially on the Reynolds
number: $\tau \propto \exp(\exp(aR+b))$~\cite{Hof:2008cba}.

Intense numerical simulations of the cPf have also been conducted, but did
not clarify the situation.  In~\cite{Faist04:JFM213501, PhysRevLett.98.014501}
a diverging behavior of the turbulent lifetimes compatible with the experimental
results of~\cite{Peixinho:2006kma,Darbyshire:2006hsa} is reported. 
Later in~\cite{Schneider:2008ig}, the authors --one of which is common
to~\cite{Faist04:JFM213501}-- conducted further simulations and reanalyzed older data,
concluding to an exponential dependence such as the one reported in~\cite{Hof:2006fk}. 
Altogether despite intense experimental and numerical effort, no definitive
answer regarding the divergence or finiteness of turbulence lifetime could be obtained
from the fit of data by phenomenological functional forms.

\subsection{Fitting procedures: lessons from glass physics}

As stated in the introduction, this issue is not specific to the transition to
turbulence. When a liquid is suddenly quenched below its crystallization temperature
and if crystallization can be avoided, the liquid enters a state, called supercooled liquid,
in which the relaxation time increases by several orders of magnitude over
a limited range of temperature~\cite{struik1978physical}. 
A divergence at a finite temperature of the relaxation time would signal an ideal
glass transition, and would thus be of high interest, at least at a conceptual level.
Despite huge efforts made to measure the variations of the relaxation time over an
experimental window of more than ten decades, no clear consensus has been
obtained yet. More precisely, the available data are both consistent with fits
including a divergence at a finite temperature $T_c>0$, and with fits diverging
only at $T=0$~\cite{PhysRevE.53.751}.

The same difficulty is also expected to appear in the context of turbulence.
We illustrate this point on experimental data recently obtained
in the case of the TCf~\cite{PhysRevE.81.025301}, when only the external
cylinder is rotating.
The TCf is then, like the pCf, linearly stable for all R. Also, because the TCf is
a closed flow, one can record very long times. In this experiment, the
angular rotation of the external cylinders fixes the Reynolds number. The flow
was perturbed by rapidly accelerating the inner cylinder in the direction
opposite to the rotation of the outer cylinder and immediately stopping it. 
After a short regime of featureless turbulence, the flow exhibits long transients 
characterized by the coexistence of laminar and turbulent domains, before
eventually relaxing towards the laminar flow. The distribution of lifetimes
is again exponential, and the authors argue that the mean turbulent lifetime
does not diverge and rather behaves in the transitional regime as a double
exponential $\tau \propto \exp(\exp(aR+b))$, as observed in the cPf~\cite{Hof:2008cba}.

\begin{figure}[t!] 
\center
\vspace{-0.5cm}
\includegraphics[width = 0.8\columnwidth,trim = 0mm 0mm 0mm 0mm, clip]{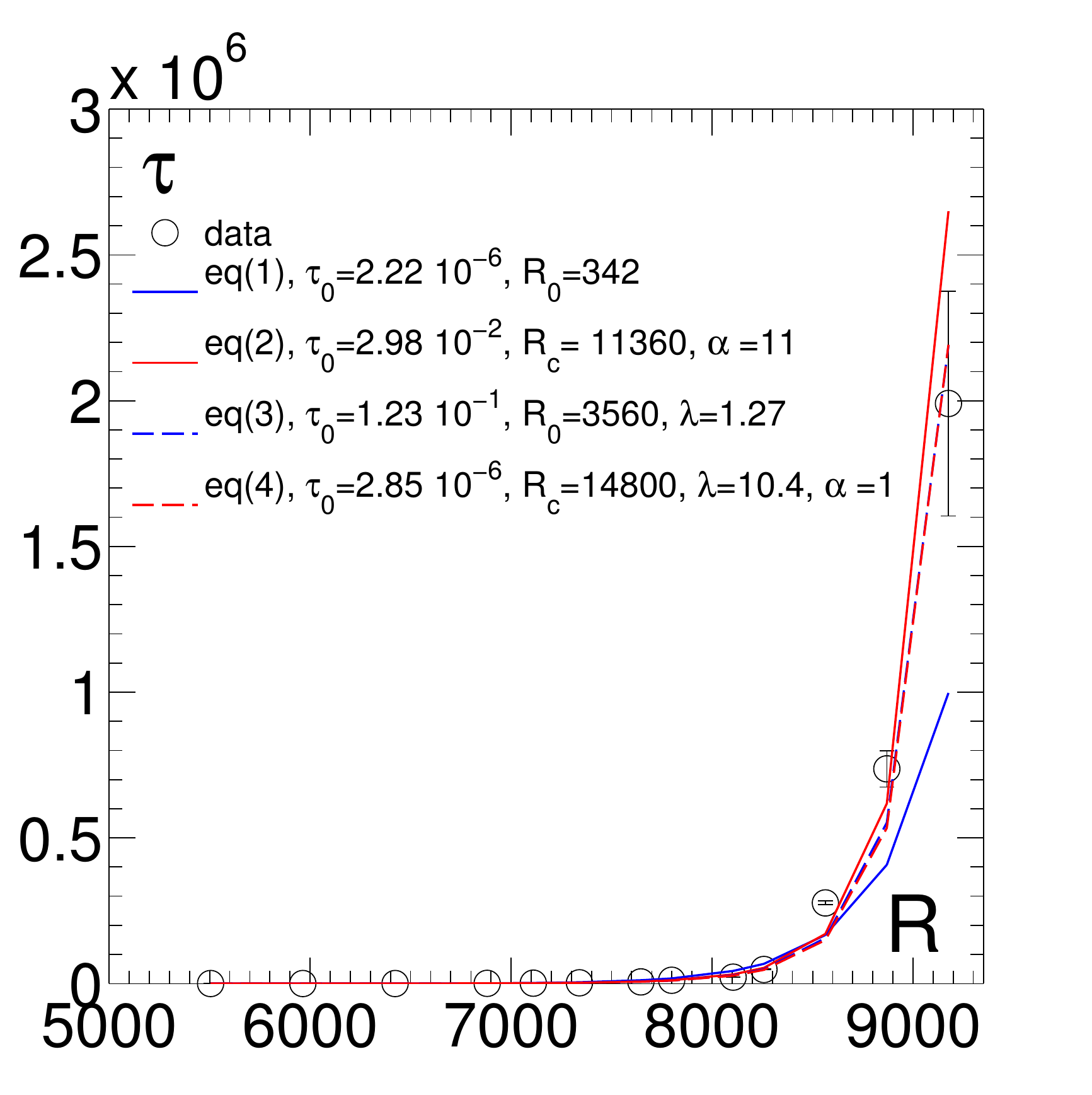}
\includegraphics[width = 0.8\columnwidth,trim = 0mm 0mm 0mm 0mm, clip]{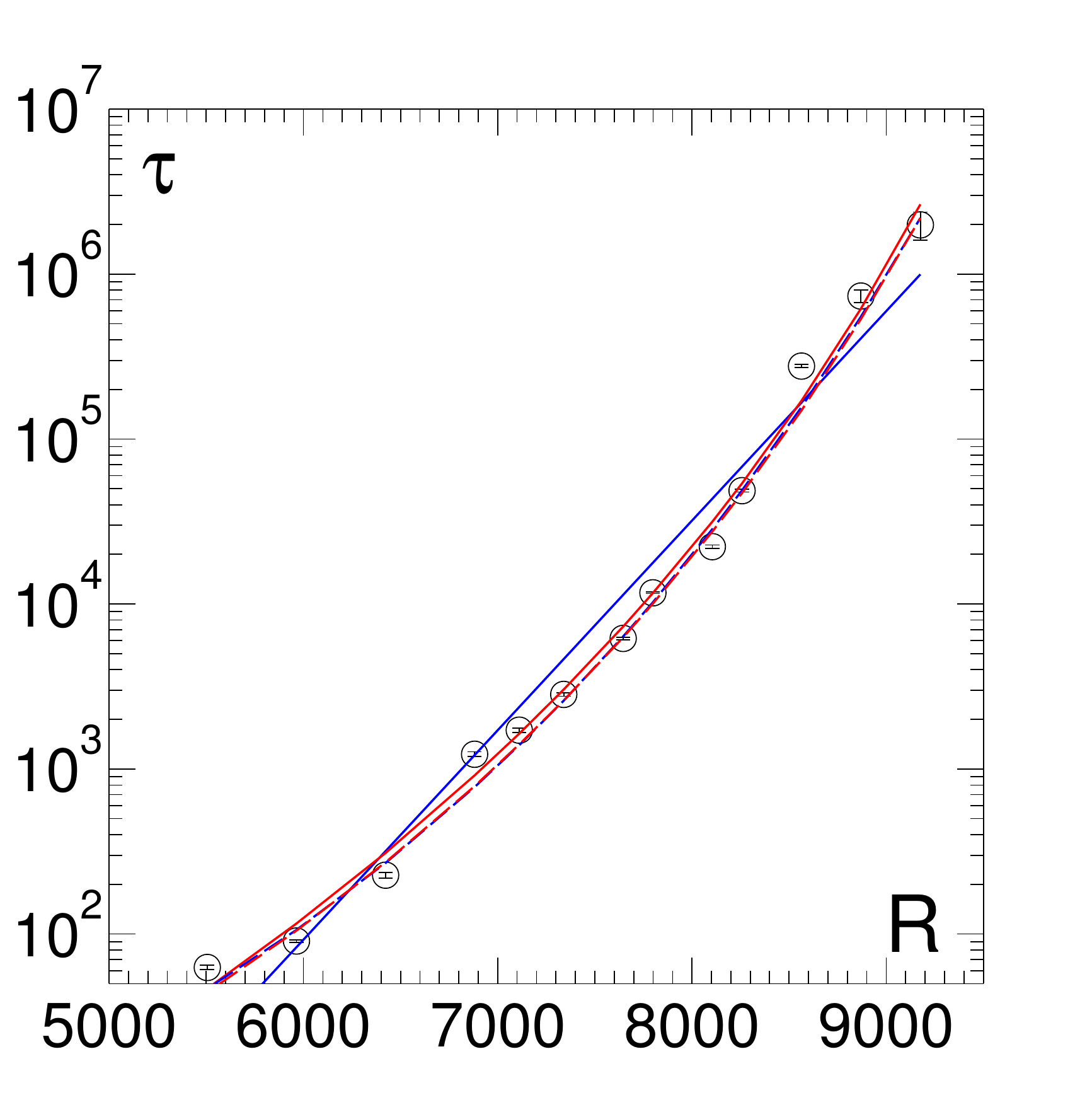}
\caption{Probing finite lifetime experimentally (color online): Relaxation lifetimes of
turbulent initial conditions in a Taylor-Couette flow, with rotating external cylinder
and internal cylinder at rest (data from~\cite{PhysRevE.81.025301}). Four possible fits are
proposed as indicated in legend. Top and bottom panels are in lin-lin and lin-log scales respectively.
Times are given in units of $d/r_0\omega_0$, where $d$ is the gap between the two cylinders,
$r_0$ is the radius of the external cylinder and $\omega_0$ its angular velocity. All fits
were performed using a standard least square fit procedure. Fit with Eq.~(1) (lowest curve on top panel and straight line on bottom panel) was obtained imposing a linear fit of $\log(\tau)$ vs.~$R$. Fit with Eq.~(2) (highest curve on top panel and convex curve on bottom panel) was obtained imposing a linear fit of $\log(\tau)$
vs.~$\log(R_c/(R_c-R))$, optimizing the fit quality on $R_c$. Fit with Eq.~(3) (dotted line, undistinguishable from fit with Eq (4)) was obtained imposing a linear fit on $\log(\tau)$ vs.~$\exp(R/R_0)$, optimizing the fit quality on $R_0$. Finally, fit with Eq.~(4) (dotted line, undistinguishable from fit with eq (3))) was obtained imposing a linear fit of $\log(\tau)$ vs.~$R_c/(R_c-R)$, optimizing the fit quality on $R_c$. In this last case, $\alpha$ was thus set to one. It was checked that other values of $\alpha$ up to $3$ cannot be discriminated.
For Eq.~(1), the regression coefficient $r_2$ is equal to $0.971$. For all the other cases, $r_2=0.994$.}
\label{fig:Tagg}
\vspace{-0.5cm}
\end{figure}

It is interesting to note that in the oldest experiments, the debate about
the functional dependence of the average turbulent lifetime on the Reynolds
number was concentrating on the choice between the two following forms:
\bea
\label{eq:exp}
&\tau/\tau_0 =& \exp(R/R_0)\\
\label{eq:div}
&\tau/\tau_0 =& \left(\frac{R_c}{R_c-R}\right)^{\alpha}, \qquad \alpha>0,
\eea
\noindent
whereas the most recent experiments, both in the case of the
cPf~\cite{Hof:2008cba} and the TCf~\cite{PhysRevE.81.025301}, have access to much longer
experimental timescales and point at  a double exponential behavior.
This last functional form ensures a very fast increase of $\tau$ without singularity,
and could give the impression that it solves the above debate.
However as learnt from the physics of glasses, the debate has actually been shifted
towards two alternative functional forms, namely: 
\bea
\label{eq:expexp}
&\ln(\tau/\tau_0) =& \lambda \exp(R/R_0)\\
\label{eq:expdiv}
&\ln(\tau/\tau_0) =& \lambda \left(\frac{R_c}{R_c-R}\right)^{\alpha}, \qquad \alpha>0.
\eea
\noindent
As a matter of fact, Eq.~(\ref{eq:expdiv}), which has (to the best of our knowledge)
not yet been proposed in the context of the transition to turbulence,
is a very standard form called the Vogel-Fulcher-Tammann
(VFT) law in the physics of glasses~\cite{PhysRevE.53.751}. 

Figure~\ref{fig:Tagg} displays the data obtained in TCf~\cite{PhysRevE.81.025301}
--which are available online as supplementary material-- together with possible fits by the
four functional forms proposed above. Note that we have only performed global
fits of the data, without trying to extract various regimes and crossovers as can
be done in the case of glasses~\cite{PhysRevE.53.751}.
One clearly observes that indeed the relevant variable to describe the growth
of the turbulent lifetimes is $\ln(\tau)$, as soon as really large times are considered.
However, one also sees that apart from the simple exponential form Eq.~(\ref{eq:exp}),
all other descriptions are not discriminable, so that there is no definitive
way to rule out or confirm the existence of a singularity. 
One faces the same difficulty as in the physics of glasses: the lifetimes to be measured
become very large, which makes it difficult  to accumulate significant statistics.
The experimental results are thereby confined to a finite range of Reynolds number
or temperature, from which even with high-quality data sets, the divergence of a
characteristic time cannot convincingly be determined from fits. 
 
Before concluding this section, let us mention that the double-exponential form Eq.~(\ref{eq:expexp})
has been justified on the basis of extreme value statistics~\cite{PhysRevE.81.035304}.
However, as stated by the authors, the argument is only local, as it involves
an expansion in $R$ around a given reference value. Hence no clear conclusion can be drawn
from the theoretical argument of~\cite{PhysRevE.81.035304} on the issue of the divergence of
$\tau$ at a finite or infinite value of $R$. Finally, let us emphasize that for now,
we have left aside all issues related to finite size effects, which in turn can severely
alter the functional dependence of time and length scales in transitional regimes. 

\section{Theoretical scenarios and models}
\label{sec-rev-dyn}

After discussing the empirical results, a natural question is to know how one can understand,
from a more theoretical perspective, the globally subcritical transition to turbulence.
This transition is by definition controlled by solutions of the
Navier-Stokes equation, which do not branch continuously from the laminar flow solution
when the Reynolds number is increased~\cite{Kawahara:2012iu}. These solutions --of various kinds, stationary
states, traveling waves, or more complex coherent structures-- are unstable and form hyperbolic
states, with stable and unstable manifolds. Early indications of the existence of these solutions
were reported in pCf, both numerically~\cite{Nagata90,CleverJFM92:397962,Cherhabili96}
and experimentally~\cite{Bottin:1998upa}. More recently, they were also observed in the cPf~\cite{deLozar:2012up}.
The intricate network made of these manifolds and their connections then serves as a skeleton for the turbulent flow. 

\subsection{Low dimensional models}

In principle one would like to collect all such states, estimate their dynamical weight and 
calculate statistical averages from periodic orbit theory~\cite{cvitanovic1991periodic}.
In practice, one must restrict the analysis to low-dimensional models
\cite{Waleffe97,Dauchot:2000uoa,Eckhardt99,moehlis2004low,Skufca:2006iu,JurgenVollmer:2009fx}
or to simulations~\cite{Schmiegel97,Itano:2001uy,Faist04:JFM213501,Skufca:2006iu,Hof:2006fk,Eckhardt:2008gd,Schneider:2007tj,Schneider:2008ig}, performed in the so-called minimal flow unit assumption~\cite{JimenezJFM91:397153}.
Doing so, it was shown that the regions of initial conditions for which long
lifetimes exhibit strong fluctuations and a sensitive dependence
on initial conditions were separated from the regions with short lifetimes and smooth variations
by a border, the so-called "edge of chaos"~\cite{Itano:2001uy,Skufca:2006iu,JurgenVollmer:2009fx}.
Later, some exact solutions with codimension-1 stable manifolds have been identified as edge states,
that is, solutions that locally form the stability boundary between laminar and turbulent
dynamics~\cite{Duguet:2009bd,Mellibovsky:2009wa,Duguet:2010dv,Schneider:2010id}.
These important results contributed to make concrete the picture borrowed from dynamical system
theory of a turbulent repellor, separated from the laminar state by a set of edge states connected
through heteroclinic manifolds. In particular the existence of the above non trivial solutions has 
served to understand the exponential distribution of lifetimes in the transitional regime.

\subsection{Spatially extended models}

Unfortunately the above picture does not bring a complete description of the subcritical transition to turbulence.
As argued in~\cite{MannevilleEPJ07,Manneville:2009if,PhysRevE.79.039904},
the reason is that the dynamics being either projected on a small set of modes or limited
to small computational domains with periodic boundary conditions, it cannot capture
the genuinely spatiotemporal coexistence of laminar and turbulent states observed in
open and unbounded flows. In particular, it can neither capture the long wavelength
modulation of turbulent intensity, nor the regime of alternating laminar and turbulent stripes,
first observed experimentally in pCf and TCf~\cite{Prigent:2002fqa,Prigent:2003uma}, and
then reproduced numerically in pCf~\cite{PhysRevLett.94.014502,Tuckerman:2008gw,Duguet:2010dv,Manneville:2010hd,MannevilleEPJ11,Manneville:2012uta}.

As a matter of fact it is known for long that, according to the scenario called spatiotemporal 
intermittency (STI~\cite{Kaneko85}), transient chaotic states locally distributed in space, e.g. on a lattice, may evolve into a sustained turbulent global state due to spatial couplings~\cite{chate1988continuous,chate1994spatiotemporal,PhysRevLett.86.5482,Grassberger91}.
Following this path, it was demonstrated that a simple 1D-model of cPf, composed of
coupled maps, does indeed captures remarkably well the character of the turbulent pipe
flow in the transitional regime and exhibits a critical transition towards sustained turbulence
via spatiotemporal intermittency~\cite{PhysRevE.84.016309}. The transition is further
believed to belong to the Directed Percolation class~\cite{chate1988continuous,chate1994spatiotemporal,grassberger2006tricritical},
as already suggested in~\cite{Pomeau86} for pCf, and recently reconsidered in cPf~\cite{sipos2011directed}.
 
Finally, it was shown by means of fully resolved direct numerical simulations
of the Navier Stokes equation, that there exists a crossover length scale of the order
of $10^2$ times the cross-stream length below which the spatio-temporal processes
at play in large-scale simulations and experiments are not captured~\cite{Manneville:2009if}.
Since then, a number of numerical investigations of large aspect ratios cPf and pCf
have reproduced the complex spatio-temporal coexistence of laminar and turbulent states,
and identified the first hydrodynamics mechanisms at play~\cite{Duguet:2010dv,Moxey:2010es,Manneville:2012uta,Avila:2011up}.

\subsection{Analogies and differences with glasses}

We now wish to discuss from a theoretical perspective the analogies, as well as the differences,
between flows close to the transition to turbulence (in short, transitional flows)
and liquids close to the glass transition.
To this aim, it is useful to first summarize the essential features of the subcritical transition to
turbulence:

\begin{itemize}
\item \emph{Subcriticality}: While the laminar flow is stable against infinitesimal perturbations,
finite amplitude perturbations may trigger an abrupt transition towards a disordered flow.
Such a disordered flow can also be obtained by quenching fully turbulent flows.
\item \emph{Spatio-temporal intermittency}: This disordered flow is made of turbulent domains, which move,
grow, decay, split and merge leading to spatio-temporal intermittency, that is a coexistence dynamics
in which active/turbulent regions may invade absorbing/laminar ones, where turbulence cannot
emerge spontaneously.
\item \emph{Transients and Meta-stability}: For large enough Reynolds number this disordered
flow has long lifetimes, which are distributed exponentially. Whether the associated characteristic time
diverges at a finite Reynolds number is still a matter of debate.  For low Reynolds number, say $R<R_u$, 
or small enough disturbances, the flow relaxes rapidly towards the laminar flow.
\item \emph{Unstable states}: When increasing the Reynolds number a larger and larger number
of unstable finite amplitude solutions appear in phase space. Some have been identified as
edge states separating the others from the laminar state.
\end{itemize}

As mentioned in the introduction, some of these features are also shared, at a qualitative level,
with glasses. For instance, the presence of long transient relaxing
states is a key feature of glasses~\cite{struik1978physical}.
Also, the existence of many unstable solutions is reminiscent
of the energy landscape picture of glasses~\cite{Sciortino:2005ks}.
Indeed, the slow relaxation in glasses has been argued to result
from the wandering of the phase-space point representing the
system in a complex energy landscape~\cite{bouchaud1997out},
mostly composed of many unstable fixed points~\cite{PhysRevLett.85.5356,PhysRevLett.85.5360,PhysRevLett.88.055502}
(though local minima also play an important role at low enough temperature).
The most striking feature of the glass transition, the rapid increase of the
relaxation time by several orders of magnitude over a moderate range of
temperature, is also interpreted as a consequence of this complex dynamics
in phase space.
These results from glass theory suggest that the complex structure
of phase space in transitional flows, with the presence of many unstable
solutions, plays an important role in the properties of the subcritical
transition to turbulence. To elaborate on this idea, we propose in
the next section an extension of the simplest model of the glass transition,
namely the Random Energy Model~\cite{PhysRevLett.45.79,PhysRevB.24.2613},
to the context of the transition to turbulence.

Other possible similarities between the transition to turbulence and the glass transition can
be pointed out, considering now the real-space dynamics.
For instance, one of the recurrent feature of glassy systems is the heterogeneities
of the dynamics: slowly and rapidly relaxing regions coexist in real space,
permanently evolving in a complex spatiotemporal organisation~\cite{vanSaarloos:2011wv}.
This is reminiscent of the dynamics observed in subcritical transitional flows,
where regions with different level of fluctuations coexist. And indeed, some
of the one-dimensional models introduced to describe such dynamical heterogeneities 
in glasses, the so-called kinetically constrained model~\cite{Chandler:2010ws, ritort2003gdk} exhibit spatiotemporal
dynamics which are very similar to those observed in the one-dimensional models
introduced to discuss the transition via spatiotemporal intermittency~\cite{chate1994spatiotemporal,PhysRevE.84.016309}, especially when looking at spatio-temporal diagrams. For some of the KCM models,
the critical point observed in the limit of zero temperature even belongs to the directed
percolation class.

Let us emphasize that beyond the possible analogies discussed above,
there are also many important differences between the glass transition
and the transition to turbulence. A first difference is that 
supercooled liquids, in which the relaxation time strongly increases
when lowering temperature close to the glass transition, are
equilibrium systems, while transitional flows are intrinsically
non-equilibrium systems. Indeed, the control parameter of the transition
(the Reynolds number) may be thought of as a distance to equilibrium,
which has to be increased to reach the turbulent state.
Another difference is that the turbulent lifetime is the time before
the flow falls into the absorbing laminar state, while the relaxation
time in glasses is defined from the relaxation of density, or stress,
correlations --no absorbing state is involved in this case.

A precise mapping between the glass transition and the transition to
turbulence should thus not be expected, and the proposed analogy should
not be considered in a strict sense.
As we shall see now, there is for instance no direct mapping between say
the Reynolds number and the temperature. 
The idea underlying the present work is rather to take advantage of the
methodological and conceptual tools developed in the framework of the
glass transition to shed some light on the subcritical transition to turbulence,
keeping in mind the limitations of such an approach.
Still, we shall see as a first illustration that it allows us to discuss
in an original way the dependence of the turbulence lifetime on
the Reynolds number.

\section{A Random Energy Model for transitional flows}
\label{sec-model}

Along the lines described in the last section, we now introduce a simple model
that captures, as an essential ingredient, the wandering of the system
on a complex landscape. This model is a variant of the
Random Energy Model~\cite{PhysRevLett.45.79,PhysRevB.24.2613},
a toy model which has proved useful in the understanding of the glass
transition, in spite of its oversimplified character.
As a byproduct, our model yields interesting predictions for the dependence
of the turbulent lifetime on the Reynolds number, as discussed below.

\subsection{Diffusion in the energy landscape}

As a first step, it is necessary to statistically characterize the properties
of the energy landscape, in particular the number of unstable solution
at a given energy above the laminar state, as function of the Reynolds
number and of the volume of the flow.
Though numerical investigations of turbulent flows have not been able yet to
characterize the number of unstable solutions as a function of the volume
of the flow, the analogy with glasses suggests that this number of solutions
may grow exponentially with the volume of the system.
Characterizing the state of the flow by its turbulent energy per unit volume,
$\ve=E/V$ (that is the excess kinetic energy with respect to the laminar flow),
we assume that the number $\Omega_V(\ve,R)$ of unstable solutions at a given
energy density $\ve$ and Reynolds number $R$ grows exponentially with the
volume $V$ according to
\be
\Omega_V(\ve,R) \sim e^{V s(\ve,R)},
\ee
thus defining an entropy density $s(\ve,R)$.
At low Reynolds number, no unstable states exist, so that we assume that the entropy
$s(\ve,R)$ is equal to zero for all $\ve>0$ if $R$ is less than a characteristic value $R_u$.
For $R > R_u$, we assume that $s(\ve,R)>0$ on an interval
$\ve_{\min}(R) < \ve < \ve_{\max}(R)$, and $s(\ve,R)=0$ otherwise,
meaning that unstable states exist only in the energy interval $(\ve_{\min}, \ve_{\max})$.

Turning to dynamics, we assume on the basis of the experimental and numerical
observations reported in section~\ref{sec-rev-dyn} that the turbulent flow spends
most of its time close to unstable solutions, and that the evolution of the flow can be
considered as a succession of jumps between different unstable solutions.
If however the flow ends up in the laminar state, the evolution stops until
an external perturbation is imposed. Taking into account the presence of the
absorbing laminar state is obviously essential to determine the lifetime of the
turbulent flow. This will be the focus of Sect.~\ref{sec-model-lifetime}.
Yet, in a first stage, it is interesting to consider the evolution of the
turbulent flow in the absence of the absorbing laminar state,
in order to make the analogy with glass models emerge more clearly.

As it is unlikely that a large amount of energy could be injected or dissipated
within a short time period, one expects that the energy of successively visited
unstable solutions are close to one another. At a coarse-grained level,
it is then natural to assume that the energy $\ve$ evolves diffusively.
In order to take into account the variation with $\ve$ of the number of unstable states,
the evolution should also be biased toward values of the energy with a high entropy $s(\ve,R)$.
More precisely, the bias should depend on the derivative of the entropy
with respect to the energy (a constant entropy introduces no bias
in the dynamics).
Altogether, the simplest evolution equation for the energy $\ve(t)$
incorporating the above ingredients is the following Langevin equation
\be \label{eq-Langevin}
\frac{d\ve}{dt} = \gamma s'(\ve,R) -\lambda + \xi(t)
\ee
where the prime denotes a derivative with respect to $\ve$.
To enforce the finite range of values $\ve_{\min} < \ve < \ve_{\max}$,
reflecting boundary conditions are assumed at $\ve_{\min}$ and $\ve_{\max}$.
The parameter $\gamma$ is a proportionality coefficient to be determined later on, 
included for dimensional reasons. The term $\lambda$ accounts for dissipative effects,
and $\xi(t)$ is a white noise describing the energy injection mechanism,
satisfying
\be \label{eq-white-noise}
\langle \xi(t) \xi(t') \rangle = 2D\, \delta(t-t'),
\ee
where $D$ is a diffusion coefficient in energy space.
These are obviously strong simplifications: the dissipation rate could in principle
depend on $\ve$ and the noise should rather be considered as colored and
multiplicative in such non-equilibrium systems, but we wish to keep the model
as simple as possible for the sake of illustration.
The assumption of a constant dissipation rate is however justified in the limit
where the width $\ve_{\max}-\ve_{\min}$ of the accessible energy range
is small with respect to $\ve_{\min}$.
Besides, considering that the noise is self-generated by the turbulent fluctuations,
and thus results from the superposition of a number of contributions
proportional to the volume $V$, one expects the diffusion coefficient to scale as
$D=D_0/V$. Note that all parameters $\gamma$, $\lambda$ and $D_0$ may
depend on the Reynolds number $R$.

\begin{figure}[t!] 
\center
\vspace{-0.0cm}
\includegraphics[width = 0.95\columnwidth,trim = 0mm 0mm 0mm 0mm, clip]{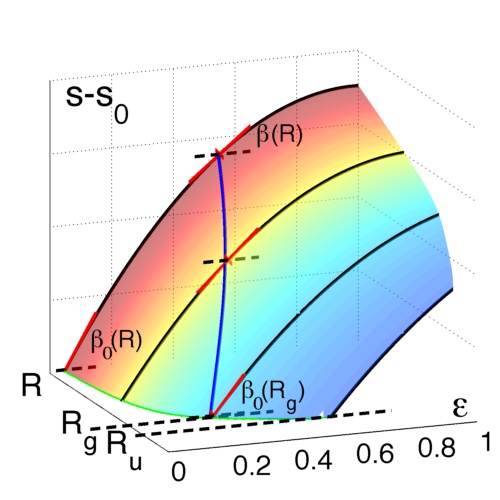}
\caption{(Color online) Sketch of the entropy surface and its slopes along the energy density direction,
together with the path followed by the flow while varying the reynolds number. Colors of
the surface go from blue to red with increasing Reynolds number $(R_g>R_u)$. The green line, on the plane $s-s_0=0$ indicates $\ve_{\min}(R)$. Fixing some Reynolds number $R$ (here a given black line amongst the four drawn on the entropy surface) one sets $\beta(R)$.
Solving Eq.~(\ref{eq-saddle}) then graphically amounts to finding a slope along the energy density
direction equal to $\beta(R)$. A solution exists if $\beta(R) < \beta_0(R)$, the slope
at the intersect with $\ve_{\min}(R)$. Varying $R$, one follows the blue path on the surface (here the line  intersects three black lines), eventually leading to the value $R_g$ such that $\beta(R_g) = \beta_0(R_g)$.}
\label{fig:entropy}
\vspace{-0.0cm}
\end{figure}

The Fokker-Planck equation describing the evolution of the probability distribution
$P(\ve,R,t)$ then reads
\be
\frac{\partial P}{\partial t} = -\frac{\partial}{\partial \ve}
\Big([\gamma s'-\lambda]\, P \Big)
+ \frac{D_0}{V} \frac{\partial^2 P}{\partial \ve^2}.
\ee
The stationary solution $P(\ve,R)$ is obtained as
\be
P(\ve,R) = \frac{1}{Z}\, \exp\left(\frac{V}{D_0} \big(\gamma s(\ve,R)-\lambda \ve \big)\right).
\ee
Following standard statistical physics arguments, one expects
the distribution $P(\ve,R)$ to be proportional to the number of unstable
states $\Omega_V(\ve,R) \sim e^{V s(\ve,R)}$, which imposes $\gamma=D_0$.
Introducing the parameter $\beta=\lambda/D_0$, the stationary distribution
then reads
\be \label{eq-dist-REM}
P(\ve,R) = \frac{1}{Z}\, \exp[V(s(\ve,R)-\beta(R) \ve)]
\ee
where we have emphasized the $R$-dependence of the parameter
$\beta$, which in the present context describes the balance between the
energy injection and the dissipation, as does the inverse temperature at equilibrium.
If $V$ is large, the distribution is dominated by the energy $\bar\ve(R)$ which
maximizes the argument of the exponential, namely $s(\ve,R)-\beta(R) \ve$. 
If the maximum of $s(\ve,R)-\beta(R)\ve$ lies within the interval
$\ve_{\min}(R) < \ve < \ve_{\max}(R)$, the most probable energy is the solution of
\be \label{eq-saddle}
s'(\bar\ve(R),R)-\beta(R) = 0.
\ee
Assuming the entropy $s(\ve,R)$ to be a concave function of $\ve$ (see figure~\ref{fig:entropy}),
$s'(\ve,R)$ is a decreasing function of $\ve$, and thus has its maximum at $\ve=\ve_{\min}(R)$.\\

We now introduce the key element of the model, which we borrow from the 
Random Energy Model \cite{PhysRevLett.45.79,PhysRevB.24.2613}.
The specificity of the latter, which leads to a glass transition,
is that the entropy has a finite slope at the minimum energy.
By analogy, we thus assume that $s'(\ve,R)$ takes a finite value, denoted as
$\beta_0(R)$, when $\ve \to \ve_{\min}(R)$.

From a statistical physics point of view, the presence of a finite slope of the entropy
at the minimum energy is related to the presence of long-range interactions in the system.
Indeed, one can check that for short-range interacting systems,
the entropy has an infinite slope at the minimal energy~\cite{Stillinger:1988es}.
In the Random Energy Model, the fact that all energy levels are statistically
independent implicitly results from long-range (mean-field type) interactions.
Indeed, the Random Energy Model can be interpreted as the limit of mean-field
spin-glass models where interactions involve $p$ spins (instead of $2$ spins
for, e.g., the Ising model), when $p \to \infty$ \cite{PhysRevB.24.2613}.

In the context of the subcritical transition to turbulence, the presence of
the pressure field and of large scale unstable solutions, such as the
unstable longitudinal vortices, naturally induces such long-range correlations.
Note also that long-range correlations are well-known to be present in the fully turbulent regime, as seen for instance by the presence of non-Gaussian fluctuations in the flow \cite{bramwell1998urf,Portelli:2003jb}
The generic presence of long-range correlations in turbulent flows thus
make plausible the assumption of a finite slope of the entropy at
the minimal energy where unstable solutions exists.
Clearly, this hypothesis would need to be checked in numerical simulations
of realistic flows, which is however a complicated task.
We thus presently take this assumption as a working hypothesis motivated
by the analogy with glasses, and explore its consequences in the framework
of subcritical turbulence modeling.

Coming back to the model, we see that if $\beta(R) < \beta_0(R)$,
Eq.~(\ref{eq-saddle}) generally admits a solution $\bar\ve(R) > \ve_{\min}(R)$.
In contrast, if $\beta(R) > \beta_0(R)$, Eq.~(\ref{eq-saddle}) has no
solution, and $s(\ve,R)-\beta(R)\ve$ is maximum at $\ve=\ve_{\min}(R)$.
The probability distribution then concentrates on $\ve_{\min}(R)$.
Intuitively, one expects $\beta(R)$ to be a decreasing function of $R$
(that is, the temperature $\beta^{-1}$ characterizing the fluctuations increase
with the Reynolds number).
On the other hand, the total number of unstable states increases with
the Reynolds number, and it is thus plausible that $\beta_0(R)$ increases
(or at least remains constant) with $R$. This suggests the existence of a
Reynolds number $R_g$ such that $\beta(R_g) = \beta_0(R_g)$.
In this case, the average energy $\bar\ve(R)$ is larger than $\ve_{\min}(R)$
for $R > R_g$, while the dynamics in phase space concentrates on the
states of minimal energy for $R < R_g$.

As emphasized at the beginning of this section, these conclusions hold under
the unphysical hypothesis that no laminar state is present. However, if the paths
leading from the unstable states to the laminar one are rare enough, the flow
is likely to visit a large number of unstable states, and should thus partially equilibrate,
before ending up into the laminar state. It is then plausible that the equilibrium distribution
given in Eq.~(\ref{eq-dist-REM}) qualitatively describes this quasi-equilibrium regime.
A natural assumption is that most of the paths leading to the laminar state
are connected to unstable states close to $\ve_{\min}(R)$, the so-called edge states
in the context of turbulence. As for $R < R_g$, the average energy remains close to
$\ve_{\min}(R)$, the flow should reach the laminar state in a reasonably short time.
Conversely, for $R > R_g$, the typical energy remains well above the
threshold $\ve_{\min}(R)$, and one expects that it takes a very large time
to find the laminar state, as it implies excursions very far from the typical energy.

Hence, the Reynolds number $R_g$ appears as a transition (or crossover)
value between a regime of short turbulent lifetime and a regime of large lifetime.
Note also that the turbulent lifetime should essentially vanish below the Reynolds
value $R_u$ where unstable states cease to exist.

\subsection{Determination of the turbulent lifetime}
\label{sec-model-lifetime}
In this section, we now try to put the above arguments on a more quantitative
basis. We define the turbulent lifetime as the mean time to reach the laminar
state after a sudden quench from a higher Reynolds number value, where
turbulence is established.
This situation can be modeled using Eq.~(\ref{eq-Langevin}) for the stochastic
dynamics of $\ve(t)$, with now an absorbing (instead of reflecting)
boundary at $\ve=\ve_{\min}$ to account for the presence of the laminar state.
The initial condition at $t=0$ is chosen as $\ve(0)=\ve_{\max}$, to model the
quench from high energy turbulent states. Determining the turbulent lifetime then amounts
to computing the mean first passage time at the absorbing boundary $\ve=\ve_{\min}$.

Such a calculation is however difficult for an arbitrary functional form of the entropy $s(\ve,R)$
and we have to restrict the choice of $s(\ve,R)$ to the linear form
\be
s(\ve,R) = \beta_0(R)\, \Big(\ve-\ve_{\min}(R) \Big) + s_0(R)
\ee
over the interval $\ve_{\min}(R) < \ve < \ve_{\max}(R)$. In this case, the mean
first passage time can be computed from the solution of the associated Fokker-Planck
equation~\cite{redner2001guide}, and one finds
\be \label{eq-tau}
\tau = \frac{V}{D_0}\, (\Delta\ve)^2 \, f\Big( V(\beta_g-\beta)\, \Delta\ve \Big)
\ee
with $\Delta \ve = \ve_{\max}-\ve_{\min}$ and $\beta_g=\beta(R_g)$, and where the function $f(x)$ is given by
\be
f(x) = \frac{1}{x^2}\, \left( e^x - 1 -x \right) \, .
\ee
For large $V$, the argument of the function $f$ in Eq.~(\ref{eq-tau}) is large as soon as
$\beta_g \ne \beta$, that is $R \ne R_g$. The value of $\tau$ is then given, to a good
approximation, by the asymptotic behavior of $f(x)$ when $x \to \pm \infty$, which reads
\bea
f(x) &\sim& \frac{1}{|x|} \qquad x \to -\infty \, ,\\
f(x) &\sim& \frac{e^{x}}{x^2} \qquad x \to +\infty \, .
\label{eq-fx-largex}
\eea
Hence, $\tau$ is given for $\beta_g< \beta$ by
\be \label{eq-tau-less}
\tau \sim \frac{\Delta \ve}{D_0 (\beta-\beta_g)} \, ,
\ee
which turns out to be independent of the volume $V$, as intuitively expected in the large $V$ limit.
In terms of Reynolds number, one thus has a power-law divergence for $R$
close to $R_g$ ($R < R_g$),
\be \label{power-law-div}
\tau \sim \frac{\tau_0}{R_g-R} \, .
\ee
However, for any finite volume $V$ this divergence is cut off when $R$ approaches
$R_g$, as soon as $R_g-R \lesssim a V^{-1}$ with some constant $a$, and a crossover
is observed to the exponential form obtained from Eq.~(\ref{eq-fx-largex})
\be \label{eq-tau-more}
\tau \sim \frac{e^{V(\beta_g-\beta) \Delta \ve}}{D_0 V (\beta_g-\beta)^2} \, . 
\ee
Contrary to Eq.~(\ref{eq-tau-less}), expression (\ref{eq-tau-more})
involves the volume $V$. For $V \to \infty$, $\tau$ becomes infinite,
and a true power-law divergence is observed for $R < R_g$.
For very large but finite $V$, the divergence can be observed in practice only on a narrow range of
Reynolds number, before $\tau$ becomes exceedingly large.
On this narrow range, $(\beta_g-\beta) \Delta \ve$ behaves linearly with $R$.
In contrast, if $V$ is not too large, the range of $R$ over which the divergence
is observed broadens, and corrections to the linear behavior of $(\beta_0-\beta) \Delta \ve$ with $R$
can become observable, possibly leading to a super-exponential behavior of $\tau$ as a function of $R$.
Though sub-exponential behavior cannot be discarded, one expects at least $\Delta \ve$
to increase with $R$, which goes in favor of the super-exponential case.

\begin{figure}[t!] 
\center
\vspace{-0.0cm}
\includegraphics[width = 0.95\columnwidth,trim = 0mm 0mm 0mm 0mm, clip]{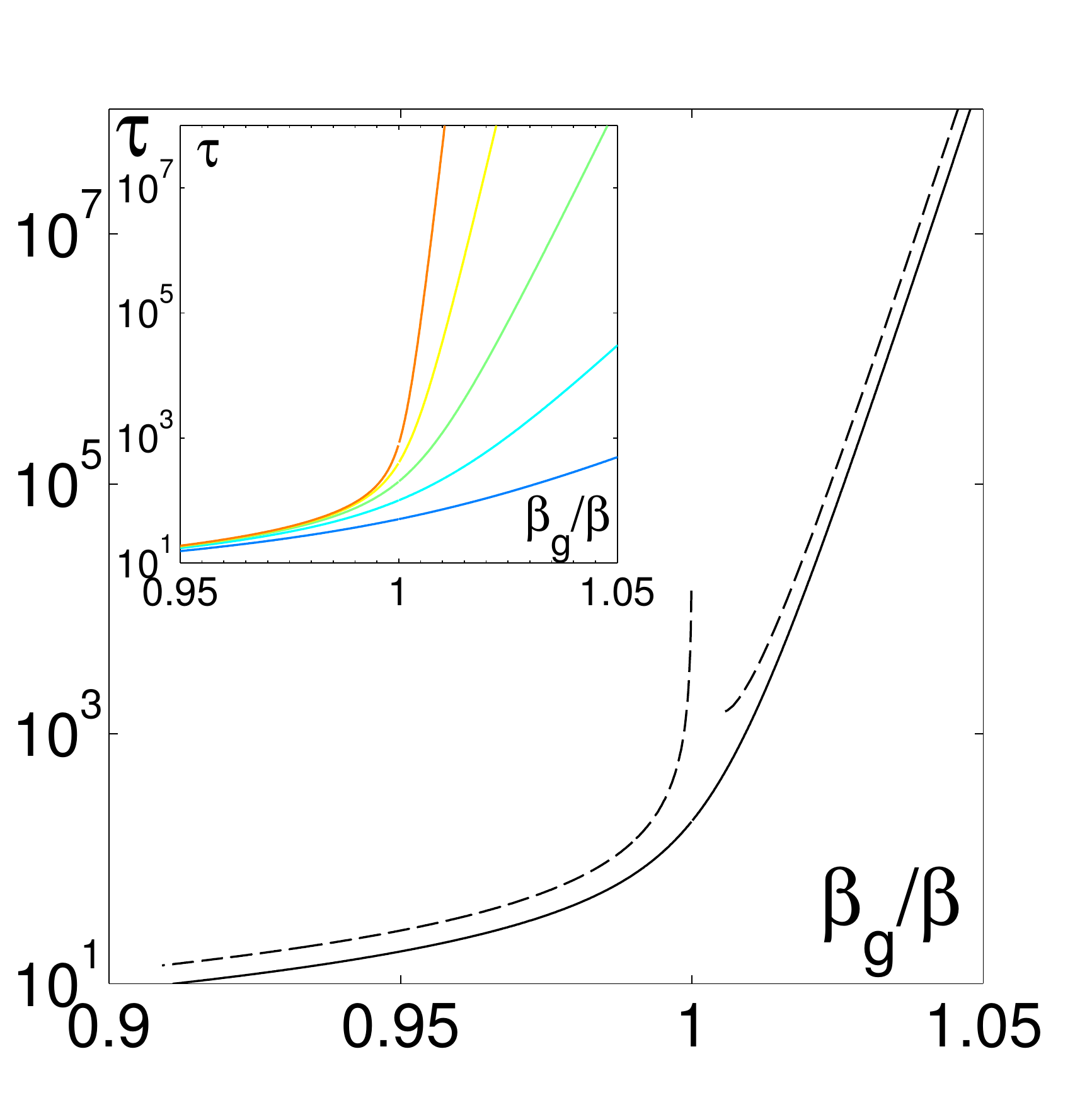}
\caption{(Color online) Sketch of the turbulent lifetime as a function of $\beta_g/\beta$, the effective Reynolds number.
Main panel: the continuous curve is the lifetime for a given volume of the system as given by Eq.~(\ref{eq-tau}); dashed curves are the asymptotic functional forms which govern the behavior of $\tau$ on each side of $\beta_g$ -they have been shifted for clarity. Inset: Turbulent lifetimes for increasing system size: the larger the system, the steeper is the increase of lifetime. The singular behavior is observed in the limit of infinite system size only.}
\label{fig:lifetimes}
\vspace{-0.0cm}
\end{figure}

\section{Discussion}

The initial motivation of the analogy proposed in this paper was two-fold.
First the intense debate that animated the transition to turbulence community
regarding the possible divergence of the turbulent lifetimes at a finite Reynolds
number  was reminiscent of a similar situation encountered in the physics of
glasses a few decades earlier. Second, the idea that the dynamics is controlled by
unstable solutions away from the laminar state shared some similarity with
the role played by the large number of saddles at the onset of the glass
transition. The goal of the analogy presented here was to make these
intuitions more precise.

We have shown that indeed, even with very good data, one cannot
discriminate a singular dependence from a regular but very fast increase
of the turbulent lifetimes, especially if one includes the possibility of a 
Vogel-Fulcher-Tammann like singularity. We have also seen that finite size
effects may lead to a crossover, which cannot be resolved experimentally
or numerically because of the extremely large timescales at play.

The model presented here was designed to be as simple as possible,
taking inspiration from the Random Energy Model with the aim
to illustrate the analogy between glasses and transitional flows.
As such, it does not claim to be realistic in any way, and some of its main
limitations are rather obvious: the spatial structure of the flow is not
taken into account, and the key ingredient (the finite slope of the
entropy at minimum energy) is taken as a working hypothesis, motivated by the analogy with glasses.
It is however quite remarkable that such a simplified model yields a crossover
between a power law and an exponential form, in qualitative agreement with
the experimental results. Note also that this result on the turbulent lifetime is not a
straightforward mapping from the Random Energy Model, since
the latter is a purely static model, with no dynamics involved,
and without any equivalent of the laminar state considered here.

These encouraging first results call for checks in direct numerical
simulations of the hypotheses underlying the model.
Counting the number of unstable solutions as a function of their
energy density, that is accessing $s(\ve,R)$, would be a major step towards
the characterization of the transition to turbulence.
This is obviously a difficult task, but still far less ambitious than characterizing the stability
properties of these solutions and describing the complex interplay of their
stable and unstable manifold. This simplification is in essence the
gain obtained when switching from a dynamical system point of view
to a statistical physics one.
A first step would be to investigate a similar approach in simpler non
linear spatio-differential equations, where spatiotemporal intermittency
has been studied, like the Kuramoto-Sivashinsky equation or the complex
Ginzburg Landau one \cite{Kaneko85}.
Valuable insights could also be obtained by measuring
in direct numerical simulations the dissipation rate as a function of the energy density,
as well as characterizing the statistical properties of the turbulent energy
fluctuations in the intermediate range of Reynolds numbers.

In the above section, we have considered $V$ as the volume of the system.
However, in the spirit of real-space approaches, the relevant volume to be
considered may rather be the volume of coherent regions of the flow, namely
regions over which correlations extend. In a very large aspect ratio experiment,
it is plausible (though not obvious) that far away regions in the system experience
no interactions. As a result, the volume $V$ would acquire a more intrinsic
nature: it would then be self-determined by the flow dynamics and not by 
the arbitrary size of the experiments.

Such a coherence volume cannot be accessed in the framework of models similar to
the Random Energy Model, which is mean-field in nature.
However, if the analogy with the physics of glasses proves to be fruitful,
it would be of interest to consider its most recent developments
(including in particular the Random First Order Transition scenario~\cite{Stevenson2005}),
which precisely address the real-space description issue~\cite{berthier2011theoretical}.
Pomeau~\cite{Pomeau86} suggested more than twenty-five years ago that the growth and death of the
laminar and turbulent regions could obey a first-order nucleation-like dynamics
(albeit of a peculiar type, given the fluctuating active property of the turbulent state
and the absorbing character of the laminar state). Let us conclude with the somewhat naive
suggestion that taking inspiration from the Random First Order Transition theory of glasses
might be a way to extend the standard laminar-turbulent coexistence scenario to
a situation where a large number of turbulent states (associated to local unstable solutions
of the Navier-Stokes equation) coexist with the laminar state.

\bibliography{/Users/olivierdauchot/Documents/_Science/Biblio/Turbulence,/Users/olivierdauchot/Documents/_Science/Biblio/Glasses}

\begin{thebibliography}{89}
\expandafter\ifx\csname natexlab\endcsname\relax\def\natexlab#1{#1}\fi
\expandafter\ifx\csname bibnamefont\endcsname\relax
  \def\bibnamefont#1{#1}\fi
\expandafter\ifx\csname bibfnamefont\endcsname\relax
  \def\bibfnamefont#1{#1}\fi
\expandafter\ifx\csname citenamefont\endcsname\relax
  \def\citenamefont#1{#1}\fi
\expandafter\ifx\csname url\endcsname\relax
  \def\url#1{\texttt{#1}}\fi
\expandafter\ifx\csname urlprefix\endcsname\relax\def\urlprefix{URL }\fi
\providecommand{\bibinfo}[2]{#2}
\providecommand{\eprint}[2][]{\url{#2}}

\bibitem[{\citenamefont{Frisch}(1995)}]{Frisch:318247}
\bibinfo{author}{\bibfnamefont{U.}~\bibnamefont{Frisch}},
  \emph{\bibinfo{title}{{Turbulence: the legacy of A N Kolmogorov}}}
  (\bibinfo{publisher}{Cambridge Univ. Press}, \bibinfo{address}{Cambridge},
  \bibinfo{year}{1995}).

\bibitem[{\citenamefont{Joseph}(1976)}]{Joseph:1976twa}
\bibinfo{author}{\bibfnamefont{D.~D.} \bibnamefont{Joseph}},
  \emph{\bibinfo{title}{{Stability of fluid motions}}}
  (\bibinfo{publisher}{Springer}, \bibinfo{year}{1976}).

\bibitem[{\citenamefont{Grossmann}(2000)}]{Grossmann:2000zz}
\bibinfo{author}{\bibfnamefont{S.}~\bibnamefont{Grossmann}},
  \bibinfo{journal}{Rev. Mod. Phys.} \textbf{\bibinfo{volume}{72}},
  \bibinfo{pages}{603} (\bibinfo{year}{2000}).

\bibitem[{\citenamefont{Dauchot and Manneville}(1997)}]{Dauchot:1997wta}
\bibinfo{author}{\bibfnamefont{O.}~\bibnamefont{Dauchot}} \bibnamefont{and}
  \bibinfo{author}{\bibfnamefont{P.}~\bibnamefont{Manneville}},
  \bibinfo{journal}{Journal de Physique II} \textbf{\bibinfo{volume}{7}},
  \bibinfo{pages}{371} (\bibinfo{year}{1997}).

\bibitem[{\citenamefont{Romanov}(1973)}]{Romanov73}
\bibinfo{author}{\bibfnamefont{V.~A.} \bibnamefont{Romanov}},
  \bibinfo{journal}{Funkcional Anal. i Prolozen} \textbf{\bibinfo{volume}{7}},
  \bibinfo{pages}{137} (\bibinfo{year}{1973}).

\bibitem[{\citenamefont{Darbyshire and Mullin}(2006)}]{Darbyshire:2006hsa}
\bibinfo{author}{\bibfnamefont{A.~G.} \bibnamefont{Darbyshire}}
  \bibnamefont{and} \bibinfo{author}{\bibfnamefont{T.}~\bibnamefont{Mullin}},
  \bibinfo{journal}{J. Fluid Mech.} \textbf{\bibinfo{volume}{289}},
  \bibinfo{pages}{83} (\bibinfo{year}{2006}).

\bibitem[{\citenamefont{Willis et~al.}(2008)\citenamefont{Willis, Peixinho,
  Kerswell, and Mullin}}]{Peixinho:2008vb}
\bibinfo{author}{\bibfnamefont{A.~P.} \bibnamefont{Willis}},
  \bibinfo{author}{\bibfnamefont{J.}~\bibnamefont{Peixinho}},
  \bibinfo{author}{\bibfnamefont{R.~R.} \bibnamefont{Kerswell}},
  \bibnamefont{and} \bibinfo{author}{\bibfnamefont{T.}~\bibnamefont{Mullin}},
  in \emph{\bibinfo{booktitle}{Philos T R Soc A}} (\bibinfo{organization}{Univ
  Bristol, Sch Math, Bristol BS8 1TW, Avon, England}, \bibinfo{year}{2008}),
  pp. \bibinfo{pages}{2671--2684}.

\bibitem[{\citenamefont{Tillmark and Alfredsson}(1992)}]{Tillmark92}
\bibinfo{author}{\bibfnamefont{N.}~\bibnamefont{Tillmark}} \bibnamefont{and}
  \bibinfo{author}{\bibfnamefont{P.~H.} \bibnamefont{Alfredsson}},
  \bibinfo{journal}{J. Fluid Mech.} \textbf{\bibinfo{volume}{235}},
  \bibinfo{pages}{89} (\bibinfo{year}{1992}).

\bibitem[{\citenamefont{Daviaud et~al.}(1992)\citenamefont{Daviaud, Hegseth,
  and Berg{\'e}}}]{PhysRevLett.69.2511}
\bibinfo{author}{\bibfnamefont{F.}~\bibnamefont{Daviaud}},
  \bibinfo{author}{\bibfnamefont{J.}~\bibnamefont{Hegseth}}, \bibnamefont{and}
  \bibinfo{author}{\bibfnamefont{P.}~\bibnamefont{Berg{\'e}}},
  \bibinfo{journal}{Phys. Rev. Lett.} \textbf{\bibinfo{volume}{69}},
  \bibinfo{pages}{2511} (\bibinfo{year}{1992}).

\bibitem[{\citenamefont{Dauchot and Daviaud}(1995)}]{Dauchot:1995tha}
\bibinfo{author}{\bibfnamefont{O.}~\bibnamefont{Dauchot}} \bibnamefont{and}
  \bibinfo{author}{\bibfnamefont{F.}~\bibnamefont{Daviaud}},
  \bibinfo{journal}{Phys. Fluids} \textbf{\bibinfo{volume}{7}},
  \bibinfo{pages}{335} (\bibinfo{year}{1995}).

\bibitem[{\citenamefont{Carlson et~al.}(1982)\citenamefont{Carlson, Widnall,
  and Peeters}}]{Carlson82}
\bibinfo{author}{\bibfnamefont{D.~R.} \bibnamefont{Carlson}},
  \bibinfo{author}{\bibfnamefont{S.~E.} \bibnamefont{Widnall}},
  \bibnamefont{and} \bibinfo{author}{\bibfnamefont{M.~F.}
  \bibnamefont{Peeters}}, \bibinfo{journal}{J. Fluid Mech.}
  \textbf{\bibinfo{volume}{{121}}}, \bibinfo{pages}{487}
  (\bibinfo{year}{1982}).

\bibitem[{\citenamefont{Coles}(1965)}]{Coles65}
\bibinfo{author}{\bibfnamefont{D.}~\bibnamefont{Coles}}, \bibinfo{journal}{J.
  Fluid Mech.} \textbf{\bibinfo{volume}{{21}}}, \bibinfo{pages}{385}
  (\bibinfo{year}{1965}).

\bibitem[{\citenamefont{Coles and Van~Atta}(1967)}]{Coles67}
\bibinfo{author}{\bibfnamefont{D.}~\bibnamefont{Coles}} \bibnamefont{and}
  \bibinfo{author}{\bibfnamefont{C.~W.} \bibnamefont{Van~Atta}},
  \bibinfo{journal}{Phys. Fluids} \textbf{\bibinfo{volume}{{10}}},
  \bibinfo{pages}{120} (\bibinfo{year}{1967}).

\bibitem[{\citenamefont{Hof et~al.}(2006)\citenamefont{Hof, Westerweel,
  Schneider, and Eckhardt}}]{Hof:2006fk}
\bibinfo{author}{\bibfnamefont{B.}~\bibnamefont{Hof}},
  \bibinfo{author}{\bibfnamefont{J.}~\bibnamefont{Westerweel}},
  \bibinfo{author}{\bibfnamefont{T.~M.} \bibnamefont{Schneider}},
  \bibnamefont{and} \bibinfo{author}{\bibfnamefont{B.}~\bibnamefont{Eckhardt}},
  \bibinfo{journal}{Nature} \textbf{\bibinfo{volume}{443}}, \bibinfo{pages}{59}
  (\bibinfo{year}{2006}).

\bibitem[{\citenamefont{Schneider
  et~al.}(2010{\natexlab{a}})\citenamefont{Schneider, Lillo, Buehrle, Eckhardt,
  D{\"o}rnemann, D{\"o}rnemann, and Freisleben}}]{TobiasMSchneider:2010gv}
\bibinfo{author}{\bibfnamefont{T.~M.} \bibnamefont{Schneider}},
  \bibinfo{author}{\bibfnamefont{F.~D.} \bibnamefont{Lillo}},
  \bibinfo{author}{\bibfnamefont{J.}~\bibnamefont{Buehrle}},
  \bibinfo{author}{\bibfnamefont{B.}~\bibnamefont{Eckhardt}},
  \bibinfo{author}{\bibfnamefont{T.}~\bibnamefont{D{\"o}rnemann}},
  \bibinfo{author}{\bibfnamefont{K.}~\bibnamefont{D{\"o}rnemann}},
  \bibnamefont{and}
  \bibinfo{author}{\bibfnamefont{B.}~\bibnamefont{Freisleben}},
  \bibinfo{journal}{Phys. Rev. E} \textbf{\bibinfo{volume}{81}}
  (\bibinfo{year}{2010}{\natexlab{a}}).

\bibitem[{\citenamefont{Faisst and Eckhardt}(2004)}]{Faist04:JFM213501}
\bibinfo{author}{\bibfnamefont{H.}~\bibnamefont{Faisst}} \bibnamefont{and}
  \bibinfo{author}{\bibfnamefont{B.}~\bibnamefont{Eckhardt}},
  \bibinfo{journal}{Journal of Fluid Mechanics} \textbf{\bibinfo{volume}{504}},
  \bibinfo{pages}{343} (\bibinfo{year}{2004}).

\bibitem[{\citenamefont{Peixinho and Mullin}(2006)}]{Peixinho:2006kma}
\bibinfo{author}{\bibfnamefont{J.}~\bibnamefont{Peixinho}} \bibnamefont{and}
  \bibinfo{author}{\bibfnamefont{T.}~\bibnamefont{Mullin}},
  \bibinfo{journal}{Phys. Rev. Lett.} \textbf{\bibinfo{volume}{96}},
  \bibinfo{pages}{094501} (\bibinfo{year}{2006}).

\bibitem[{\citenamefont{Bottin et~al.}(1998{\natexlab{a}})\citenamefont{Bottin,
  Daviaud, Manneville, and Dauchot}}]{Bottin:1998ve}
\bibinfo{author}{\bibfnamefont{S.}~\bibnamefont{Bottin}},
  \bibinfo{author}{\bibfnamefont{F.}~\bibnamefont{Daviaud}},
  \bibinfo{author}{\bibfnamefont{P.}~\bibnamefont{Manneville}},
  \bibnamefont{and} \bibinfo{author}{\bibfnamefont{O.}~\bibnamefont{Dauchot}},
  \bibinfo{journal}{Europhys. Lett.} \textbf{\bibinfo{volume}{43}},
  \bibinfo{pages}{171} (\bibinfo{year}{1998}{\natexlab{a}}).

\bibitem[{\citenamefont{Bottin and Chat{\'e}}(1998)}]{bottinEPJ98}
\bibinfo{author}{\bibfnamefont{S.}~\bibnamefont{Bottin}} \bibnamefont{and}
  \bibinfo{author}{\bibfnamefont{H.}~\bibnamefont{Chat{\'e}}},
  \bibinfo{journal}{The European Physical Journal B - Condensed Matter and
  Complex Systems} \textbf{\bibinfo{volume}{6}}, \bibinfo{pages}{143}
  (\bibinfo{year}{1998}).

\bibitem[{\citenamefont{Prigent et~al.}(2002)\citenamefont{Prigent,
  Gr{\'e}goire, Chat{\'e}, Dauchot, and van Saarloos}}]{Prigent:2002fqa}
\bibinfo{author}{\bibfnamefont{A.}~\bibnamefont{Prigent}},
  \bibinfo{author}{\bibfnamefont{G.}~\bibnamefont{Gr{\'e}goire}},
  \bibinfo{author}{\bibfnamefont{H.}~\bibnamefont{Chat{\'e}}},
  \bibinfo{author}{\bibfnamefont{O.}~\bibnamefont{Dauchot}}, \bibnamefont{and}
  \bibinfo{author}{\bibfnamefont{W.}~\bibnamefont{van Saarloos}},
  \bibinfo{journal}{Phys. Rev. Lett.} \textbf{\bibinfo{volume}{89}},
  \bibinfo{pages}{014501} (\bibinfo{year}{2002}).

\bibitem[{\citenamefont{Prigent et~al.}(2003)\citenamefont{Prigent,
  Gr{\'e}goire, Chat{\'e}, and Dauchot}}]{Prigent:2003uma}
\bibinfo{author}{\bibfnamefont{A.}~\bibnamefont{Prigent}},
  \bibinfo{author}{\bibfnamefont{G.}~\bibnamefont{Gr{\'e}goire}},
  \bibinfo{author}{\bibfnamefont{H.}~\bibnamefont{Chat{\'e}}},
  \bibnamefont{and} \bibinfo{author}{\bibfnamefont{O.}~\bibnamefont{Dauchot}},
  \bibinfo{journal}{Physica D: Nonlinear Phenomena}
  \textbf{\bibinfo{volume}{174}}, \bibinfo{pages}{100} (\bibinfo{year}{2003}).

\bibitem[{\citenamefont{Borrero-Echeverry
  et~al.}(2010)\citenamefont{Borrero-Echeverry, Schatz, and
  Tagg}}]{PhysRevE.81.025301}
\bibinfo{author}{\bibfnamefont{D.}~\bibnamefont{Borrero-Echeverry}},
  \bibinfo{author}{\bibfnamefont{M.~F.} \bibnamefont{Schatz}},
  \bibnamefont{and} \bibinfo{author}{\bibfnamefont{R.}~\bibnamefont{Tagg}},
  \bibinfo{journal}{Phys. Rev. E} \textbf{\bibinfo{volume}{81}},
  \bibinfo{pages}{025301} (\bibinfo{year}{2010}).

\bibitem[{\citenamefont{Avila et~al.}(2011)\citenamefont{Avila, Moxey,
  de~Lozar, Avila, Barkley, and Hof}}]{Avila:2011up}
\bibinfo{author}{\bibfnamefont{K.}~\bibnamefont{Avila}},
  \bibinfo{author}{\bibfnamefont{D.}~\bibnamefont{Moxey}},
  \bibinfo{author}{\bibfnamefont{A.}~\bibnamefont{de~Lozar}},
  \bibinfo{author}{\bibfnamefont{M.}~\bibnamefont{Avila}},
  \bibinfo{author}{\bibfnamefont{D.}~\bibnamefont{Barkley}}, \bibnamefont{and}
  \bibinfo{author}{\bibfnamefont{B.}~\bibnamefont{Hof}},
  \bibinfo{journal}{Science} \textbf{\bibinfo{volume}{333}},
  \bibinfo{pages}{192} (\bibinfo{year}{2011}).

\bibitem[{\citenamefont{Samanta et~al.}(2011)\citenamefont{Samanta, De~Lozar,
  and Hof}}]{Samanta:2011va}
\bibinfo{author}{\bibfnamefont{D.}~\bibnamefont{Samanta}},
  \bibinfo{author}{\bibfnamefont{A.}~\bibnamefont{De~Lozar}}, \bibnamefont{and}
  \bibinfo{author}{\bibfnamefont{B.}~\bibnamefont{Hof}}, \bibinfo{journal}{J.
  Fluid Mech.} \textbf{\bibinfo{volume}{1}}, \bibinfo{pages}{1}
  (\bibinfo{year}{2011}).

\bibitem[{\citenamefont{Eckhardt et~al.}(2008)\citenamefont{Eckhardt, Faisst,
  Schmiegel, and Schneider}}]{Eckhardt:2008gd}
\bibinfo{author}{\bibfnamefont{B.}~\bibnamefont{Eckhardt}},
  \bibinfo{author}{\bibfnamefont{H.}~\bibnamefont{Faisst}},
  \bibinfo{author}{\bibfnamefont{A.}~\bibnamefont{Schmiegel}},
  \bibnamefont{and} \bibinfo{author}{\bibfnamefont{T.~M.}
  \bibnamefont{Schneider}}, \bibinfo{journal}{Philosophical Transactions of the
  Royal Society A: Mathematical, Physical and Engineering Sciences}
  \textbf{\bibinfo{volume}{366}}, \bibinfo{pages}{1297} (\bibinfo{year}{2008}).

\bibitem[{\citenamefont{Lagha and Manneville}(2007)}]{MannevilleEPJ07}
\bibinfo{author}{\bibfnamefont{M.}~\bibnamefont{Lagha}} \bibnamefont{and}
  \bibinfo{author}{\bibfnamefont{P.}~\bibnamefont{Manneville}},
  \bibinfo{journal}{Eur. Phys. J. B} \textbf{\bibinfo{volume}{58}},
  \bibinfo{pages}{433} (\bibinfo{year}{2007}).

\bibitem[{\citenamefont{Gibson et~al.}(2008)\citenamefont{Gibson, Halcrow, and
  Cvitanovic}}]{Gibson:2008ec}
\bibinfo{author}{\bibfnamefont{J.~F.} \bibnamefont{Gibson}},
  \bibinfo{author}{\bibfnamefont{J.}~\bibnamefont{Halcrow}}, \bibnamefont{and}
  \bibinfo{author}{\bibfnamefont{P.}~\bibnamefont{Cvitanovic}},
  \bibinfo{journal}{J. Fluid Mech.} \textbf{\bibinfo{volume}{611}},
  \bibinfo{pages}{107} (\bibinfo{year}{2008}).

\bibitem[{\citenamefont{Manneville}(2009{\natexlab{a}})}]{Manneville:2009if}
\bibinfo{author}{\bibfnamefont{P.}~\bibnamefont{Manneville}},
  \bibinfo{journal}{Phys. Rev. E} \textbf{\bibinfo{volume}{79}},
  \bibinfo{pages}{025301(R)} (\bibinfo{year}{2009}{\natexlab{a}}).

\bibitem[{\citenamefont{Barkley}(2011)}]{PhysRevE.84.016309}
\bibinfo{author}{\bibfnamefont{D.}~\bibnamefont{Barkley}},
  \bibinfo{journal}{Phys. Rev. E} \textbf{\bibinfo{volume}{84}},
  \bibinfo{pages}{016309} (\bibinfo{year}{2011}).

\bibitem[{\citenamefont{Barkley and Tuckerman}(2005)}]{PhysRevLett.94.014502}
\bibinfo{author}{\bibfnamefont{D.}~\bibnamefont{Barkley}} \bibnamefont{and}
  \bibinfo{author}{\bibfnamefont{L.~S.} \bibnamefont{Tuckerman}},
  \bibinfo{journal}{Phys. Rev. Lett.} \textbf{\bibinfo{volume}{94}},
  \bibinfo{pages}{014502} (\bibinfo{year}{2005}).

\bibitem[{\citenamefont{Willis and Kerswell}(2007)}]{PhysRevLett.98.014501}
\bibinfo{author}{\bibfnamefont{A.~P.} \bibnamefont{Willis}} \bibnamefont{and}
  \bibinfo{author}{\bibfnamefont{R.~R.} \bibnamefont{Kerswell}},
  \bibinfo{journal}{Phys. Rev. Lett.} \textbf{\bibinfo{volume}{98}},
  \bibinfo{pages}{014501} (\bibinfo{year}{2007}).

\bibitem[{\citenamefont{Schneider and Eckhardt}(2008)}]{Schneider:2008ig}
\bibinfo{author}{\bibfnamefont{T.~M.} \bibnamefont{Schneider}}
  \bibnamefont{and} \bibinfo{author}{\bibfnamefont{B.}~\bibnamefont{Eckhardt}},
  \bibinfo{journal}{Phys. Rev. E} \textbf{\bibinfo{volume}{78}}
  (\bibinfo{year}{2008}).

\bibitem[{\citenamefont{Duguet et~al.}(2008)\citenamefont{Duguet, Willis, and
  Kerswell}}]{Duguet:2008bs}
\bibinfo{author}{\bibfnamefont{Y.}~\bibnamefont{Duguet}},
  \bibinfo{author}{\bibfnamefont{A.~P.} \bibnamefont{Willis}},
  \bibnamefont{and} \bibinfo{author}{\bibfnamefont{R.~R.}
  \bibnamefont{Kerswell}}, \bibinfo{journal}{J. Fluid Mech.}
  \textbf{\bibinfo{volume}{613}}, \bibinfo{pages}{255} (\bibinfo{year}{2008}).

\bibitem[{\citenamefont{Tuckerman et~al.}(2008)\citenamefont{Tuckerman,
  Barkley, and Dauchot}}]{Tuckerman:2008gw}
\bibinfo{author}{\bibfnamefont{L.~S.} \bibnamefont{Tuckerman}},
  \bibinfo{author}{\bibfnamefont{D.}~\bibnamefont{Barkley}}, \bibnamefont{and}
  \bibinfo{author}{\bibfnamefont{O.}~\bibnamefont{Dauchot}},
  \bibinfo{journal}{J. Phys.: Conf. Ser.} \textbf{\bibinfo{volume}{137}},
  \bibinfo{pages}{012029} (\bibinfo{year}{2008}).

\bibitem[{\citenamefont{Duguet et~al.}(2009)\citenamefont{Duguet, Schlatter,
  and Henningson}}]{Duguet:2009bd}
\bibinfo{author}{\bibfnamefont{Y.}~\bibnamefont{Duguet}},
  \bibinfo{author}{\bibfnamefont{P.}~\bibnamefont{Schlatter}},
  \bibnamefont{and} \bibinfo{author}{\bibfnamefont{D.~S.}
  \bibnamefont{Henningson}}, \bibinfo{journal}{Phys. Fluids}
  \textbf{\bibinfo{volume}{21}}, \bibinfo{pages}{111701}
  (\bibinfo{year}{2009}).

\bibitem[{\citenamefont{Schneider
  et~al.}(2010{\natexlab{b}})\citenamefont{Schneider, Marinc, and
  Eckhardt}}]{Schneider:2010id}
\bibinfo{author}{\bibfnamefont{T.~M.} \bibnamefont{Schneider}},
  \bibinfo{author}{\bibfnamefont{D.}~\bibnamefont{Marinc}}, \bibnamefont{and}
  \bibinfo{author}{\bibfnamefont{B.}~\bibnamefont{Eckhardt}},
  \bibinfo{journal}{J. Fluid Mech.} \textbf{\bibinfo{volume}{646}},
  \bibinfo{pages}{441} (\bibinfo{year}{2010}{\natexlab{b}}).

\bibitem[{\citenamefont{Duguet et~al.}(2010)\citenamefont{Duguet, Schlatter,
  and Henningson}}]{Duguet:2010dv}
\bibinfo{author}{\bibfnamefont{Y.}~\bibnamefont{Duguet}},
  \bibinfo{author}{\bibfnamefont{P.}~\bibnamefont{Schlatter}},
  \bibnamefont{and} \bibinfo{author}{\bibfnamefont{D.~S.}
  \bibnamefont{Henningson}}, \bibinfo{journal}{J. Fluid Mech.}
  \textbf{\bibinfo{volume}{650}}, \bibinfo{pages}{119} (\bibinfo{year}{2010}).

\bibitem[{\citenamefont{Moxey and Barkley}(2010)}]{Moxey:2010es}
\bibinfo{author}{\bibfnamefont{D.}~\bibnamefont{Moxey}} \bibnamefont{and}
  \bibinfo{author}{\bibfnamefont{D.}~\bibnamefont{Barkley}},
  \bibinfo{journal}{Proceedings of the National Academy of Sciences}
  \textbf{\bibinfo{volume}{107}}, \bibinfo{pages}{8091} (\bibinfo{year}{2010}).

\bibitem[{\citenamefont{Manneville and Rolland}(2011)}]{Manneville:2010hd}
\bibinfo{author}{\bibfnamefont{P.}~\bibnamefont{Manneville}} \bibnamefont{and}
  \bibinfo{author}{\bibfnamefont{J.}~\bibnamefont{Rolland}},
  \bibinfo{journal}{Theor. Comput. Fluid. Dyn.} \textbf{\bibinfo{volume}{25}},
  \bibinfo{pages}{407} (\bibinfo{year}{2011}).

\bibitem[{\citenamefont{Philip and Manneville}(2011)}]{Philip:2011ja}
\bibinfo{author}{\bibfnamefont{J.}~\bibnamefont{Philip}} \bibnamefont{and}
  \bibinfo{author}{\bibfnamefont{P.}~\bibnamefont{Manneville}},
  \bibinfo{journal}{Phys. Rev. E} \textbf{\bibinfo{volume}{83}},
  \bibinfo{pages}{036308} (\bibinfo{year}{2011}).

\bibitem[{\citenamefont{Manneville}(2012)}]{Manneville:2012uta}
\bibinfo{author}{\bibfnamefont{P.}~\bibnamefont{Manneville}},
  \bibinfo{journal}{arXiv} \textbf{\bibinfo{volume}{physics.flu-dyn}}
  (\bibinfo{year}{2012}).

\bibitem[{\citenamefont{Struik}(1978)}]{struik1978physical}
\bibinfo{author}{\bibfnamefont{L.}~\bibnamefont{Struik}},
  \emph{\bibinfo{title}{{Physical aging in amorphous polymers and other
  materials}}} (\bibinfo{publisher}{Elsevier Amsterdam}, \bibinfo{year}{1978}).

\bibitem[{\citenamefont{Angelani et~al.}(2000)\citenamefont{Angelani,
  Di~Leonardo, Ruocco, Scala, and Sciortino}}]{PhysRevLett.85.5356}
\bibinfo{author}{\bibfnamefont{L.}~\bibnamefont{Angelani}},
  \bibinfo{author}{\bibfnamefont{R.}~\bibnamefont{Di~Leonardo}},
  \bibinfo{author}{\bibfnamefont{G.}~\bibnamefont{Ruocco}},
  \bibinfo{author}{\bibfnamefont{A.}~\bibnamefont{Scala}}, \bibnamefont{and}
  \bibinfo{author}{\bibfnamefont{F.}~\bibnamefont{Sciortino}},
  \bibinfo{journal}{Phys. Rev. Lett.} \textbf{\bibinfo{volume}{85}},
  \bibinfo{pages}{5356} (\bibinfo{year}{2000}).

\bibitem[{\citenamefont{Berthier and Biroli}(2011)}]{berthier2011theoretical}
\bibinfo{author}{\bibfnamefont{L.}~\bibnamefont{Berthier}} \bibnamefont{and}
  \bibinfo{author}{\bibfnamefont{G.}~\bibnamefont{Biroli}},
  \bibinfo{journal}{Review of Modern Physics} \textbf{\bibinfo{volume}{83}},
  \bibinfo{pages}{587} (\bibinfo{year}{2011}).

\bibitem[{\citenamefont{Kivelson et~al.}(1996)\citenamefont{Kivelson, Tarjus,
  Zhao, and Kivelson}}]{PhysRevE.53.751}
\bibinfo{author}{\bibfnamefont{D.}~\bibnamefont{Kivelson}},
  \bibinfo{author}{\bibfnamefont{G.}~\bibnamefont{Tarjus}},
  \bibinfo{author}{\bibfnamefont{X.}~\bibnamefont{Zhao}}, \bibnamefont{and}
  \bibinfo{author}{\bibfnamefont{S.~A.} \bibnamefont{Kivelson}},
  \bibinfo{journal}{Phys. Rev. E} \textbf{\bibinfo{volume}{53}},
  \bibinfo{pages}{751} (\bibinfo{year}{1996}).

\bibitem[{\citenamefont{Derrida}(1980)}]{PhysRevLett.45.79}
\bibinfo{author}{\bibfnamefont{B.}~\bibnamefont{Derrida}},
  \bibinfo{journal}{Phys. Rev. Lett.} \textbf{\bibinfo{volume}{45}},
  \bibinfo{pages}{79} (\bibinfo{year}{1980}).

\bibitem[{\citenamefont{Derrida}(1981)}]{PhysRevB.24.2613}
\bibinfo{author}{\bibfnamefont{B.}~\bibnamefont{Derrida}},
  \bibinfo{journal}{Phys. Rev. B} \textbf{\bibinfo{volume}{24}},
  \bibinfo{pages}{2613} (\bibinfo{year}{1981}).

\bibitem[{\citenamefont{Pomeau}(1986)}]{Pomeau86}
\bibinfo{author}{\bibfnamefont{Y.}~\bibnamefont{Pomeau}},
  \bibinfo{journal}{Physica D} \textbf{\bibinfo{volume}{23}},
  \bibinfo{pages}{3} (\bibinfo{year}{1986}).

\bibitem[{\citenamefont{Hof et~al.}(2008)\citenamefont{Hof, de~Lozar, Kuik, and
  Westerweel}}]{Hof:2008cba}
\bibinfo{author}{\bibfnamefont{B.}~\bibnamefont{Hof}},
  \bibinfo{author}{\bibfnamefont{A.}~\bibnamefont{de~Lozar}},
  \bibinfo{author}{\bibfnamefont{D.}~\bibnamefont{Kuik}}, \bibnamefont{and}
  \bibinfo{author}{\bibfnamefont{J.}~\bibnamefont{Westerweel}},
  \bibinfo{journal}{Phys. Rev. Lett.} \textbf{\bibinfo{volume}{101}},
  \bibinfo{pages}{214501} (\bibinfo{year}{2008}).

\bibitem[{\citenamefont{Goldenfeld et~al.}(2010)\citenamefont{Goldenfeld,
  Guttenberg, and Gioia}}]{PhysRevE.81.035304}
\bibinfo{author}{\bibfnamefont{N.}~\bibnamefont{Goldenfeld}},
  \bibinfo{author}{\bibfnamefont{N.}~\bibnamefont{Guttenberg}},
  \bibnamefont{and} \bibinfo{author}{\bibfnamefont{G.}~\bibnamefont{Gioia}},
  \bibinfo{journal}{Phys. Rev. E} \textbf{\bibinfo{volume}{81}},
  \bibinfo{pages}{035304} (\bibinfo{year}{2010}).

\bibitem[{\citenamefont{Kawahara et~al.}(2012)\citenamefont{Kawahara, Uhlmann,
  and van Veen}}]{Kawahara:2012iu}
\bibinfo{author}{\bibfnamefont{G.}~\bibnamefont{Kawahara}},
  \bibinfo{author}{\bibfnamefont{M.}~\bibnamefont{Uhlmann}}, \bibnamefont{and}
  \bibinfo{author}{\bibfnamefont{L.}~\bibnamefont{van Veen}},
  \bibinfo{journal}{Annual Review of Fluid Mechanics}
  \textbf{\bibinfo{volume}{44}}, \bibinfo{pages}{203} (\bibinfo{year}{2012}).

\bibitem[{\citenamefont{Nagata}(1990)}]{Nagata90}
\bibinfo{author}{\bibfnamefont{M.}~\bibnamefont{Nagata}}, \bibinfo{journal}{J.
  Fluid Mech.} \textbf{\bibinfo{volume}{217}}, \bibinfo{pages}{519}
  (\bibinfo{year}{1990}).

\bibitem[{\citenamefont{Clever and Busse}(1992)}]{CleverJFM92:397962}
\bibinfo{author}{\bibfnamefont{R.~M.} \bibnamefont{Clever}} \bibnamefont{and}
  \bibinfo{author}{\bibfnamefont{F.~H.} \bibnamefont{Busse}},
  \bibinfo{journal}{Journal of Fluid Mechanics} \textbf{\bibinfo{volume}{234}},
  \bibinfo{pages}{511} (\bibinfo{year}{1992}).

\bibitem[{\citenamefont{Cherhabili and Ehrenstein}(1996)}]{Cherhabili96}
\bibinfo{author}{\bibfnamefont{A.}~\bibnamefont{Cherhabili}} \bibnamefont{and}
  \bibinfo{author}{\bibfnamefont{U.}~\bibnamefont{Ehrenstein}},
  \bibinfo{journal}{J. Fluid Mech.} \textbf{\bibinfo{volume}{342}},
  \bibinfo{pages}{159} (\bibinfo{year}{1996}).

\bibitem[{\citenamefont{Bottin et~al.}(1998{\natexlab{b}})\citenamefont{Bottin,
  Dauchot, Daviaud, and Manneville}}]{Bottin:1998upa}
\bibinfo{author}{\bibfnamefont{S.}~\bibnamefont{Bottin}},
  \bibinfo{author}{\bibfnamefont{O.}~\bibnamefont{Dauchot}},
  \bibinfo{author}{\bibfnamefont{F.}~\bibnamefont{Daviaud}}, \bibnamefont{and}
  \bibinfo{author}{\bibfnamefont{P.}~\bibnamefont{Manneville}},
  \bibinfo{journal}{Phys. Fluids} \textbf{\bibinfo{volume}{10}},
  \bibinfo{pages}{2597} (\bibinfo{year}{1998}{\natexlab{b}}).

\bibitem[{\citenamefont{de~Lozar et~al.}(2012)\citenamefont{de~Lozar,
  Mellibovsky, Avila, and Hof}}]{deLozar:2012up}
\bibinfo{author}{\bibfnamefont{A.}~\bibnamefont{de~Lozar}},
  \bibinfo{author}{\bibfnamefont{F.}~\bibnamefont{Mellibovsky}},
  \bibinfo{author}{\bibfnamefont{M.}~\bibnamefont{Avila}}, \bibnamefont{and}
  \bibinfo{author}{\bibfnamefont{B.}~\bibnamefont{Hof}},
  \bibinfo{journal}{Phys. Rev. Lett.} \textbf{\bibinfo{volume}{108}}
  (\bibinfo{year}{2012}).

\bibitem[{\citenamefont{Cvitanovic and
  Eckhardt}(1991)}]{cvitanovic1991periodic}
\bibinfo{author}{\bibfnamefont{P.}~\bibnamefont{Cvitanovic}} \bibnamefont{and}
  \bibinfo{author}{\bibfnamefont{B.}~\bibnamefont{Eckhardt}},
  \bibinfo{journal}{Journal of Physics A: Mathematical and General}
  \textbf{\bibinfo{volume}{24}}, \bibinfo{pages}{L237} (\bibinfo{year}{1991}).

\bibitem[{\citenamefont{Waleffe}(1997)}]{Waleffe97}
\bibinfo{author}{\bibfnamefont{F.}~\bibnamefont{Waleffe}},
  \bibinfo{journal}{Phys. of Fluids} \textbf{\bibinfo{volume}{9}},
  \bibinfo{pages}{883} (\bibinfo{year}{1997}).

\bibitem[{\citenamefont{Dauchot and Vioujard}(2000)}]{Dauchot:2000uoa}
\bibinfo{author}{\bibfnamefont{O.}~\bibnamefont{Dauchot}} \bibnamefont{and}
  \bibinfo{author}{\bibfnamefont{N.}~\bibnamefont{Vioujard}},
  \bibinfo{journal}{The European Physical Journal B - Condensed Matter and
  Complex Systems} \textbf{\bibinfo{volume}{14}}, \bibinfo{pages}{377}
  (\bibinfo{year}{2000}).

\bibitem[{\citenamefont{Eckhardt and Mersmann}(1999)}]{Eckhardt99}
\bibinfo{author}{\bibfnamefont{B.}~\bibnamefont{Eckhardt}} \bibnamefont{and}
  \bibinfo{author}{\bibfnamefont{A.}~\bibnamefont{Mersmann}},
  \bibinfo{journal}{Phys. Rev. E} \textbf{\bibinfo{volume}{60}},
  \bibinfo{pages}{509} (\bibinfo{year}{1999}).

\bibitem[{\citenamefont{Moehlis et~al.}(2004)\citenamefont{Moehlis, Faisst, and
  Eckhardt}}]{moehlis2004low}
\bibinfo{author}{\bibfnamefont{J.}~\bibnamefont{Moehlis}},
  \bibinfo{author}{\bibfnamefont{H.}~\bibnamefont{Faisst}}, \bibnamefont{and}
  \bibinfo{author}{\bibfnamefont{B.}~\bibnamefont{Eckhardt}},
  \bibinfo{journal}{New Journal of Physics} \textbf{\bibinfo{volume}{6}},
  \bibinfo{pages}{56} (\bibinfo{year}{2004}).

\bibitem[{\citenamefont{Skufca et~al.}(2006)\citenamefont{Skufca, Yorke, and
  Eckhardt}}]{Skufca:2006iu}
\bibinfo{author}{\bibfnamefont{J.}~\bibnamefont{Skufca}},
  \bibinfo{author}{\bibfnamefont{J.}~\bibnamefont{Yorke}}, \bibnamefont{and}
  \bibinfo{author}{\bibfnamefont{B.}~\bibnamefont{Eckhardt}},
  \bibinfo{journal}{Phys. Rev. Lett.} \textbf{\bibinfo{volume}{96}},
  \bibinfo{pages}{174101} (\bibinfo{year}{2006}).

\bibitem[{\citenamefont{Vollmer et~al.}(2009)\citenamefont{Vollmer, Schneider,
  and Eckhardt}}]{JurgenVollmer:2009fx}
\bibinfo{author}{\bibfnamefont{J.}~\bibnamefont{Vollmer}},
  \bibinfo{author}{\bibfnamefont{T.~M.} \bibnamefont{Schneider}},
  \bibnamefont{and} \bibinfo{author}{\bibfnamefont{B.}~\bibnamefont{Eckhardt}},
  \bibinfo{journal}{New J. Phys.} \textbf{\bibinfo{volume}{11}}
  (\bibinfo{year}{2009}).

\bibitem[{\citenamefont{Schmiegel and Eckhardt}(1997)}]{Schmiegel97}
\bibinfo{author}{\bibfnamefont{A.}~\bibnamefont{Schmiegel}} \bibnamefont{and}
  \bibinfo{author}{\bibfnamefont{B.}~\bibnamefont{Eckhardt}},
  \bibinfo{journal}{Phys. Rev. Lett.} \textbf{\bibinfo{volume}{79}},
  \bibinfo{pages}{5250} (\bibinfo{year}{1997}).

\bibitem[{\citenamefont{Itano and Toh}(2001)}]{Itano:2001uy}
\bibinfo{author}{\bibfnamefont{T.}~\bibnamefont{Itano}} \bibnamefont{and}
  \bibinfo{author}{\bibfnamefont{S.}~\bibnamefont{Toh}}, \bibinfo{journal}{J.
  Phys. Soc. Jpn.} \textbf{\bibinfo{volume}{70}}, \bibinfo{pages}{703}
  (\bibinfo{year}{2001}).

\bibitem[{\citenamefont{Schneider et~al.}(2007)\citenamefont{Schneider,
  Eckhardt, and Yorke}}]{Schneider:2007tj}
\bibinfo{author}{\bibfnamefont{T.~M.} \bibnamefont{Schneider}},
  \bibinfo{author}{\bibfnamefont{B.}~\bibnamefont{Eckhardt}}, \bibnamefont{and}
  \bibinfo{author}{\bibfnamefont{J.~A.} \bibnamefont{Yorke}},
  \bibinfo{journal}{Phys. Rev. Lett.} \textbf{\bibinfo{volume}{99}}
  (\bibinfo{year}{2007}).

\bibitem[{\citenamefont{Jimenez and Moin}(1991)}]{JimenezJFM91:397153}
\bibinfo{author}{\bibfnamefont{J.}~\bibnamefont{Jimenez}} \bibnamefont{and}
  \bibinfo{author}{\bibfnamefont{P.}~\bibnamefont{Moin}},
  \bibinfo{journal}{Journal of Fluid Mechanics} \textbf{\bibinfo{volume}{225}},
  \bibinfo{pages}{213} (\bibinfo{year}{1991}).

\bibitem[{\citenamefont{Mellibovsky et~al.}(2009)\citenamefont{Mellibovsky,
  Meseguer, Schneider, and Eckhardt}}]{Mellibovsky:2009wa}
\bibinfo{author}{\bibfnamefont{F.}~\bibnamefont{Mellibovsky}},
  \bibinfo{author}{\bibfnamefont{A.}~\bibnamefont{Meseguer}},
  \bibinfo{author}{\bibfnamefont{T.~M.} \bibnamefont{Schneider}},
  \bibnamefont{and} \bibinfo{author}{\bibfnamefont{B.}~\bibnamefont{Eckhardt}},
  \bibinfo{journal}{Phys. Rev. Lett.} \textbf{\bibinfo{volume}{103}}
  (\bibinfo{year}{2009}).

\bibitem[{\citenamefont{Manneville}(2009{\natexlab{b}})}]{PhysRevE.79.039904}
\bibinfo{author}{\bibfnamefont{P.}~\bibnamefont{Manneville}},
  \bibinfo{journal}{Phys. Rev. E} \textbf{\bibinfo{volume}{79}},
  \bibinfo{pages}{039904} (\bibinfo{year}{2009}{\natexlab{b}}).

\bibitem[{\citenamefont{Rolland and Manneville}(2011)}]{MannevilleEPJ11}
\bibinfo{author}{\bibfnamefont{J.}~\bibnamefont{Rolland}} \bibnamefont{and}
  \bibinfo{author}{\bibfnamefont{P.}~\bibnamefont{Manneville}},
  \bibinfo{journal}{Eur. Phys. J. B} \textbf{\bibinfo{volume}{80}},
  \bibinfo{pages}{529} (\bibinfo{year}{2011}).

\bibitem[{\citenamefont{Kaneko}(1985)}]{Kaneko85}
\bibinfo{author}{\bibfnamefont{K.}~\bibnamefont{Kaneko}},
  \bibinfo{journal}{Prog. Theor. Phys.} \textbf{\bibinfo{volume}{{73}}},
  \bibinfo{pages}{1033} (\bibinfo{year}{1985}).

\bibitem[{\citenamefont{Chat{\'e} and Manneville}(1988)}]{chate1988continuous}
\bibinfo{author}{\bibfnamefont{H.}~\bibnamefont{Chat{\'e}}} \bibnamefont{and}
  \bibinfo{author}{\bibfnamefont{P.}~\bibnamefont{Manneville}},
  \bibinfo{journal}{Europhys. Lett.} \textbf{\bibinfo{volume}{6}},
  \bibinfo{pages}{591} (\bibinfo{year}{1988}).

\bibitem[{\citenamefont{Chat{\'e} and
  Manneville}(1994)}]{chate1994spatiotemporal}
\bibinfo{author}{\bibfnamefont{H.}~\bibnamefont{Chat{\'e}}} \bibnamefont{and}
  \bibinfo{author}{\bibfnamefont{P.}~\bibnamefont{Manneville}}, in
  \emph{\bibinfo{booktitle}{Turbulence: A tentative dictionnary}}, edited by
  \bibinfo{editor}{\bibfnamefont{P.}~\bibnamefont{Tabeling}} \bibnamefont{and}
  \bibinfo{editor}{\bibfnamefont{O.}~\bibnamefont{Cardoso}}
  (\bibinfo{publisher}{Plenum Press}, \bibinfo{year}{1994}), p.
  \bibinfo{pages}{111}.

\bibitem[{\citenamefont{Bohr et~al.}(2001)\citenamefont{Bohr, van Hecke,
  Mikkelsen, and Ipsen}}]{PhysRevLett.86.5482}
\bibinfo{author}{\bibfnamefont{T.}~\bibnamefont{Bohr}},
  \bibinfo{author}{\bibfnamefont{M.}~\bibnamefont{van Hecke}},
  \bibinfo{author}{\bibfnamefont{R.}~\bibnamefont{Mikkelsen}},
  \bibnamefont{and} \bibinfo{author}{\bibfnamefont{M.}~\bibnamefont{Ipsen}},
  \bibinfo{journal}{Phys. Rev. Lett.} \textbf{\bibinfo{volume}{86}},
  \bibinfo{pages}{5482} (\bibinfo{year}{2001}).

\bibitem[{\citenamefont{Grassberger and Schreiber}(1991)}]{Grassberger91}
\bibinfo{author}{\bibfnamefont{P.}~\bibnamefont{Grassberger}} \bibnamefont{and}
  \bibinfo{author}{\bibfnamefont{T.}~\bibnamefont{Schreiber}},
  \bibinfo{journal}{Physica D} \textbf{\bibinfo{volume}{{50}}},
  \bibinfo{pages}{177} (\bibinfo{year}{1991}).

\bibitem[{\citenamefont{Grassberger}(2006)}]{grassberger2006tricritical}
\bibinfo{author}{\bibfnamefont{P.}~\bibnamefont{Grassberger}},
  \bibinfo{journal}{J. Stat. Mech.} \textbf{\bibinfo{volume}{2006}},
  \bibinfo{pages}{P01004} (\bibinfo{year}{2006}).

\bibitem[{\citenamefont{Sipos and Goldenfeld}(2011)}]{sipos2011directed}
\bibinfo{author}{\bibfnamefont{M.}~\bibnamefont{Sipos}} \bibnamefont{and}
  \bibinfo{author}{\bibfnamefont{N.}~\bibnamefont{Goldenfeld}},
  \bibinfo{journal}{Phys. Rev. E} \textbf{\bibinfo{volume}{84}},
  \bibinfo{pages}{035304} (\bibinfo{year}{2011}).

\bibitem[{\citenamefont{Sciortino}(2005)}]{Sciortino:2005ks}
\bibinfo{author}{\bibfnamefont{F.}~\bibnamefont{Sciortino}},
  \bibinfo{journal}{J. Stat. Mech.} \textbf{\bibinfo{volume}{2005}},
  \bibinfo{pages}{P05015} (\bibinfo{year}{2005}).

\bibitem[{\citenamefont{Bouchaud et~al.}(1998)\citenamefont{Bouchaud,
  Cugliandolo, Kurchan, and Mezard}}]{bouchaud1997out}
\bibinfo{author}{\bibfnamefont{J.-P.} \bibnamefont{Bouchaud}},
  \bibinfo{author}{\bibfnamefont{L.}~\bibnamefont{Cugliandolo}},
  \bibinfo{author}{\bibfnamefont{J.}~\bibnamefont{Kurchan}}, \bibnamefont{and}
  \bibinfo{author}{\bibfnamefont{M.}~\bibnamefont{Mezard}}, in
  \emph{\bibinfo{booktitle}{Spin Glasses and Random Fields}}
  (\bibinfo{publisher}{Young, A.~P. World Scientific, Singapore},
  \bibinfo{year}{1998}).

\bibitem[{\citenamefont{Broderix et~al.}(2000)\citenamefont{Broderix,
  Bhattacharya, Cavagna, Zippelius, and Giardina}}]{PhysRevLett.85.5360}
\bibinfo{author}{\bibfnamefont{K.}~\bibnamefont{Broderix}},
  \bibinfo{author}{\bibfnamefont{K.~K.} \bibnamefont{Bhattacharya}},
  \bibinfo{author}{\bibfnamefont{A.}~\bibnamefont{Cavagna}},
  \bibinfo{author}{\bibfnamefont{A.}~\bibnamefont{Zippelius}},
  \bibnamefont{and} \bibinfo{author}{\bibfnamefont{I.}~\bibnamefont{Giardina}},
  \bibinfo{journal}{Phys. Rev. Lett.} \textbf{\bibinfo{volume}{85}},
  \bibinfo{pages}{5360} (\bibinfo{year}{2000}).

\bibitem[{\citenamefont{Grigera et~al.}(2002)\citenamefont{Grigera, Cavagna,
  Giardina, and Parisi}}]{PhysRevLett.88.055502}
\bibinfo{author}{\bibfnamefont{T.~S.} \bibnamefont{Grigera}},
  \bibinfo{author}{\bibfnamefont{A.}~\bibnamefont{Cavagna}},
  \bibinfo{author}{\bibfnamefont{I.}~\bibnamefont{Giardina}}, \bibnamefont{and}
  \bibinfo{author}{\bibfnamefont{G.}~\bibnamefont{Parisi}},
  \bibinfo{journal}{Phys. Rev. Lett.} \textbf{\bibinfo{volume}{88}},
  \bibinfo{pages}{055502} (\bibinfo{year}{2002}).

\bibitem[{\citenamefont{van Saarloos et~al.}(2011)\citenamefont{van Saarloos,
  Cipelletti, Bouchaud, Biroli, and Berthier}}]{vanSaarloos:2011wv}
\bibinfo{editor}{\bibfnamefont{W.}~\bibnamefont{van Saarloos}},
  \bibinfo{editor}{\bibfnamefont{L.}~\bibnamefont{Cipelletti}},
  \bibinfo{editor}{\bibfnamefont{J.-P.} \bibnamefont{Bouchaud}},
  \bibinfo{editor}{\bibfnamefont{G.}~\bibnamefont{Biroli}}, \bibnamefont{and}
  \bibinfo{editor}{\bibfnamefont{L.}~\bibnamefont{Berthier}}, eds.,
  \emph{\bibinfo{title}{{Dynamical Heterogeneities in Glasses, Colloids, and
  Granular Media}}}, International Series of Monographs on Physics 150
  (\bibinfo{publisher}{Oxford University Press}, \bibinfo{address}{Oxford},
  \bibinfo{year}{2011}).

\bibitem[{\citenamefont{Chandler and Garrahan}(2010)}]{Chandler:2010ws}
\bibinfo{author}{\bibfnamefont{D.}~\bibnamefont{Chandler}} \bibnamefont{and}
  \bibinfo{author}{\bibfnamefont{J.~P.} \bibnamefont{Garrahan}},
  \bibinfo{journal}{Annual Review of Physical Chemistry}
  \textbf{\bibinfo{volume}{61}}, \bibinfo{pages}{191} (\bibinfo{year}{2010}).

\bibitem[{\citenamefont{Ritort and Sollich}(2003)}]{ritort2003gdk}
\bibinfo{author}{\bibfnamefont{F.}~\bibnamefont{Ritort}} \bibnamefont{and}
  \bibinfo{author}{\bibfnamefont{P.}~\bibnamefont{Sollich}},
  \bibinfo{journal}{Advances in Physics} \textbf{\bibinfo{volume}{52}},
  \bibinfo{pages}{219} (\bibinfo{year}{2003}).

\bibitem[{\citenamefont{Stillinger}(1988)}]{Stillinger:1988es}
\bibinfo{author}{\bibfnamefont{F.~H.} \bibnamefont{Stillinger}},
  \bibinfo{journal}{The Journal of Chemical Physics}
  \textbf{\bibinfo{volume}{88}}, \bibinfo{pages}{7818} (\bibinfo{year}{1988}).

\bibitem[{\citenamefont{Bramwell et~al.}(1998)\citenamefont{Bramwell,
  Holdsworth, and Pinton}}]{bramwell1998urf}
\bibinfo{author}{\bibfnamefont{S.~T.} \bibnamefont{Bramwell}},
  \bibinfo{author}{\bibfnamefont{P.~C.~W.} \bibnamefont{Holdsworth}},
  \bibnamefont{and} \bibinfo{author}{\bibfnamefont{J.~F.}
  \bibnamefont{Pinton}}, \bibinfo{journal}{Nature}
  \textbf{\bibinfo{volume}{396}}, \bibinfo{pages}{552} (\bibinfo{year}{1998}).

\bibitem[{\citenamefont{Portelli et~al.}(2003)\citenamefont{Portelli,
  Holdsworth, and Pinton}}]{Portelli:2003jb}
\bibinfo{author}{\bibfnamefont{B.}~\bibnamefont{Portelli}},
  \bibinfo{author}{\bibfnamefont{P.}~\bibnamefont{Holdsworth}},
  \bibnamefont{and} \bibinfo{author}{\bibfnamefont{J.~F.}
  \bibnamefont{Pinton}}, \bibinfo{journal}{Phys. Rev. Lett.}
  \textbf{\bibinfo{volume}{90}}, \bibinfo{pages}{104501}
  (\bibinfo{year}{2003}).

\bibitem[{\citenamefont{Redner}(2001)}]{redner2001guide}
\bibinfo{author}{\bibfnamefont{S.}~\bibnamefont{Redner}},
  \emph{\bibinfo{title}{{A guide to first-passage processes}}}
  (\bibinfo{publisher}{Cambridge Univ Pr}, \bibinfo{year}{2001}).

\bibitem[{\citenamefont{Stevenson and Wolynes}(2005)}]{Stevenson2005}
\bibinfo{author}{\bibfnamefont{J.}~\bibnamefont{Stevenson}} \bibnamefont{and}
  \bibinfo{author}{\bibfnamefont{P.}~\bibnamefont{Wolynes}},
  \bibinfo{journal}{J. Phys. Chem. B} \textbf{\bibinfo{volume}{109}},
  \bibinfo{pages}{15093} (\bibinfo{year}{2005}).

\end{thebibliography}


\end{document}